\documentclass{emulateapj}

\def\be{\begin{equation}}
\def\ee{\end{equation}}
\def\ergs{\rm erg~s^{-1}}
\def\bhpr{_{\bullet,0}}
\def\I{_{P}}
\def\D{_{D}}
\def\bol{_{\rm bol}}
\def\bulge{_{\rm bulge}}
\def\gal{_{\rm gal}}
\def\Edd{_{\rm Edd}}
\def\bhtI{_{\bullet,{P}}}

\def\d{{\rm d}}
\def\early{^{\rm e}}
\def\bol{_{\rm bol}}
\def\bulge{_{\rm bulge}}
\def\gal{_{\rm gal}}
\def\Edd{_{\rm Edd}}

\def\d{{\rm d}}
\def\early{^{\rm e}}
\def\late{^{\rm l}}
\def\Mpc{{\rm\,Mpc}}
\def\c{_{\rm c}}
\def\dex{{\rm\,dex}}
\def\bh{M_{\bullet}}
\def\bhp{M_{\bullet,1}}
\def\bhs{M_{\bullet,2}}
\def\mbh{M_{\bullet,0}}

\def\life{_{\rm lt}}
\def\taus{\tau_{\rm S}}

\def\yr{\rm yr}
\def\msun{M_{\odot}}
\def\kms{{\rm km~s^{-1}}}

\shorttitle{Precise Constraints on MBH Growth}
\shortauthors{Yu \& Lu }

\begin{document}
\title{Toward precise constraints on growth of massive black holes }
\author{Qingjuan Yu\altaffilmark{1,3} and Youjun Lu\altaffilmark{2,1,3} }
\altaffiltext{1}{Canadian Institute for Theoretical Astrophysics and Department of Astronomy and Astrophysics, Toronto, ON M5S 3H8, Canada; Email: yuqj,luyj@cita.utoronto.ca}
\altaffiltext{2}{National Astronomical Observatories, Chinese Academy of Sciences, Beijing, 100012, China}
\altaffiltext{3}{Department of Astronomy and Astrophysics, University of California, Santa Cruz, CA 95064}

\begin{abstract}

Growth of massive black holes (MBHs) in galactic centers comes mainly from gas
accretion during their QSO/AGN phases. In this paper we apply an extended
So{\l}tan argument, connecting the local MBH mass function with the
time-integral of the QSO luminosity function, to the demography of MBHs and
QSOs from recent optical and X-ray surveys, and obtain robust constraints on
the luminosity evolution (or mass growth history) of individual QSOs (or MBHs).
We find that the luminosity evolution probably involves two phases: an initial
exponentially increasing phase set by the Eddington limit and a following phase
in which the luminosity declines with time as a power law (with a slope of
$\sim -1.2$---$-1.3$) set by a self-similar long-term evolution of disk
accretion. Neither an evolution involving only the increasing phase with a
single Eddington ratio nor an exponentially declining pattern in the second
phase is likely. The period of a QSO radiating at a luminosity higher than 10\%
of its peak value is about 2--3$\times10^8\yr$, during which the MBH obtains
$\sim 80\%$ of its mass. The mass-to-energy conversion efficiency is
$\simeq0.16\pm0.04 ^{+0.05}_{-0}$, with the latter error accounting for the
maximum uncertainty due to Compton-thick AGNs. The expected Eddington ratios in
QSOs from the constrained luminosity evolution cluster around a single
value close to 0.5--1 for high-luminosity QSOs and extend to a wide range of
lower values for low-luminosity ones. The Eddington ratios for high luminosity
QSOs appear to conflict with those estimated from observations ($\sim 0.25$) by
using some virial mass estimators for MBHs in QSOs unless the estimators
systematically over-estimate MBH masses by a factor of 2--4. We also infer the
fraction of optically obscured QSOs $\sim 60-80\%$.
The constraints obtained above are not affected significantly by
MBH mergers and multiple-times of nuclear activity (e.g., triggered by
multiple times of galaxy wet major mergers) in the MBH growth history.
We discuss further
applications of the luminosity evolution of individual QSOs to obtaining the
MBH mass function at high redshifts and the cosmic evolution of triggering
rates of nuclear activity.

\end{abstract}
\maketitle

\section{Introduction}\label{sec:intro}

Massive black holes (MBHs), probably remnants of QSOs \citep{LyndenBell}, have
been detected in the nuclei of many nearby galaxies \citep{KR95, Magorrian98,
Richstone, KG01, FF05}. How do these local MBHs form and evolve, and what is
the most important mechanism shaping the mass distribution of MBHs? The current
consensus is that the local MBHs obtained their mass mainly through accretion
during phases of nuclear activity when they appeared as
QSOs/AGNs,\footnote{Hereafter, we frequently use the term QSOs rather than
QSOs/AGNs, if not otherwise specified, to represent QSOs and/or AGNs
for convenience.} similar to the ones seen now in the distant universe
\citep[e.g.,][]{YT02,YL04a, Marconi04, Shankar04, Barger05, Hopkins06,
Shankar07}.  The evolution of mass accretion onto a MBH is equivalent to the
luminosity evolution, given the mass-to-energy conversion efficiency, and is
recorded in the luminosity function (LF) of QSOs.  However, the QSO LF depends
mainly on two functions: (1) ${\cal G}(z;\mbh)$, the rate of nuclear activity
triggered at different redshifts $z$ for MBHs with present-day mass $\mbh$; (2)
${\cal L}(\tau;\mbh)$, the luminosity evolution history of a QSO, of which the
remnant MBH has a present-day mass $\mbh$, as a function of the age of its
nuclear activity $\tau$.  One cannot derive these two functions only from the
knowledge of the QSO LF without additional assumptions.

In an extended version of the \citet{S82} argument, the local MBH mass
distribution function (BHMF) is related to QSOs found in the distant universe
by the simple integral equation 
\begin{eqnarray} 
\int^{\infty}_0
\Psi_{L}(L,z)\left|\frac{dt}{dz}\right|dz & = & \int^{\infty}_{0}
n_{\bh}(\mbh,t_0) \times \nonumber \\ &  & \tau\life(\mbh)P(L|\mbh) d\mbh,
\label{eq:relation} 
\end{eqnarray} 
where $t_0$ is the present cosmic time, $n_{\bh}(\mbh,t_0)$ is the local BHMF,
defined so that $n_{\bh}(\mbh,t_0)d\mbh$ gives the number density of local MBHs
with present-day mass in the range $\mbh\rightarrow \mbh+d\mbh$,
$\Psi_{L}(L,z)$ is the QSO LF, defined so that $\Psi_{L}(L,z)dL$ gives the
comoving number density of QSOs with nuclear luminosity in the range
$L\rightarrow L+dL$ at redshift $z$,
\be
\tau\life(\mbh)=\int dL\sum_{k} \frac{1}{\left|\frac{d{\cal
L}(\tau;\mbh)}{d\tau}|_{\tau=\tau_k(L,\mbh)}\right|} 
\label{eq:taulife} 
\ee 
is the time interval (or the QSO lifetime) in which that a MBH with present-day
mass $\mbh$ appeared as a QSO, and $\tau_k(L,\mbh)$ $(k=1,2,...)$ are the roots
of the equation ${\cal L}(\tau;\mbh)-L=0$ (see details of the derivation in
\citealt{YL04a}).  Here ${\cal L}(\tau;\mbh)$ represents the luminosity of a
QSO and its associated MBH with present-day mass $\mbh$ at a time $\tau$ after
the triggering of nuclear activity.  The value of $\tau\life$ depends on the
detailed definition of ``active nuclei'' or the lower threshold set to the
nuclear luminosity. Finally,
\be
P(L|\mbh)=\frac{1}{\tau\life(\mbh)}\sum_{k} \frac{1}{\left|\frac{d{\cal
L}(\tau;\mbh)}{d\tau}|_{\tau=\tau_k(L,\mbh)}\right|} 
\label{eq:PLMbh} 
\ee 
is the probability distribution function of the nuclear (bolometric) luminosity
$L$ over the growth history of the MBH.  The right-hand-side of equation
(\ref{eq:relation}) gives the total time spent per unit $L$ at luminosity $L$
by the progenitors of all the local MBHs in a unit comoving volume, which
should be the time integral of the QSO LF, i.e., the left-hand-side of the
equation.  Multiplying equation (\ref{eq:relation}) by the BH mass accretion
rate $(1-\epsilon)L/(\epsilon c^2)=\dot\bh$ (see eqs.~\ref{eq:mdotinf} and
\ref{eq:mdot} below), where $\epsilon$ is the mass-to-energy conversion
efficiency and $c$ is the speed of light, and then integrating it over cosmic
time $t$ reduces to the So{\l}tan (1982) argument \citep{YL04a}.  Provided that
two basic quantities, i.e., the local BHMF and the QSO LF, can be
observationally determined with sufficient accuracy, the kernel
$\tau\life(\mbh)P(L|\mbh)$, containing information on the luminosity evolution
history of individual QSOs/MBHs,  may be solved from the integral
equation~(\ref{eq:relation}). Therefore, the extended So{\l}tan argument is
expected to give robust but more detailed constraints on the growth of MBHs
than the simple energetic argument due to \citet{S82}. 

As an alternative approach to the theoretical models based on the hierarchical
co-evolution of MBHs and galaxies/galactic halos studied intensively in the
literature \citep[e.g.,][]{ER88, HR93, HNR98, KH00, Granato01, WL03,
Volonteri03, Croton06, Bower06,Malbon07}, in this paper we use the integral
equation (\ref{eq:relation}) to statistically constrain the growth history of
individual MBHs or ${\cal L}$. The advantages of this approach are: (1) the
accretion history of individual QSOs, ${\cal L}(\tau;\mbh)$, is isolated from
the triggering rate of nuclear activity, ${\cal G}(z;\mbh)$, which is
presumably associated with mergers of galaxies or instabilities of galactic
disks; and (2) it is free of the many adjustable parameters introduced in the
co-evolution models and probably also avoids uncertain assumptions on seed BHs.
Note that these two functions,  ${\cal G}(z;\mbh)$ and ${\cal L}(\tau;\mbh)$,
are mixed in the differential continuity equation for BHMF evolution presented
in \citet[][see also \citealt{Cavaliere71}, \citealt{CP89}, and
\citealt{CP90}]{SB92}, which is widely used in studying the growth of MBHs
\citep[e.g.,][]{Marconi04, Shankar07}. Using the luminosity evolution curves,
i.e., ${\cal L}(\tau;\mbh)$, obtained from numerical simulations of colliding
galaxies, \citet{Hopkins06} elaborated a unified model for the origin of QSOs
and MBHs (see also their other papers listed therein). A possible concern with
that approach is that simulations of colliding galaxies have a spatial
resolution much larger than the scale of accretion disks around MBHs and
therefore may not reflect the real luminosity evolution, as the disk accretion
is probably self-regulated in the vicinity of MBHs rather than being directly
determined by the material infall rate from a much larger scale or the
Bondi-accretion rate (see discussions in \S~\ref{sec:models}). (For another
model of the possible light curve, see \citealt{CO07}.)

Estimating the local BHMF can be done with recent advances in observations
\citep[e.g.,][]{Salucci99,AR02,YL04a,Marconi04,Shankar04,Lauer07a,Tundo07}.
First, MBHs are believed to exist in the nuclei of most, if not all, nearby
galaxies \citep{KR95,Magorrian98,Richstone,KG01,FF05}. Second, it has been well
established that tight correlations exist between the MBH mass and various
galactic properties, such as mass, luminosity, stellar velocity-dispersion,
light concentration and binding energy of the hot components of galaxies
\citep[here hot components mean either ellipticals or spiral bulges;][]{KR95,
Magorrian98, FM00, Gebhardt00, Tremaine02, HR04, MH04, Graham01, AR07}.  Third,
the luminosity or velocity-dispersion functions of nearby galaxies have been
well determined by large surveys such as the Sloan Digital Sky Survey
\citep[SDSS;][]{Blanton03, Bernardi03, Sheth03}.  Combining the correlation
between the MBH mass and galaxy velocity dispersion (or luminosity) with the
velocity-dispersion (or luminosity) distribution of nearby galaxies, we
estimate the local BHMF in \S~\ref{sec:BHMF}. 

In the past several years, the QSO LF has been determined over unprecedentedly
large luminosity and redshift ranges both from optical surveys such as the Two
Degree Field QSO Redshift Survey (2Qz) and SDSS, and from X-ray surveys by
ASCA, Chandra and XMM-Newton. For example, the optical QSO LF has been obtained
over the redshift range $0.4<z<2.1$ and the magnitude range $M_{\rm b_J}<-22.5$
using a sample of more than 15,000 QSOs from 2Qz \citep{Croom04};
\citet{Richards06a} estimated the QSO LF over a larger redshift range
($0.3<z<5$), but only for bright QSOs, using a homogeneous statistical sample
of 15,343 QSOs drawn from SDSS Data Release 3; using the COMBO-17 data,
\citet{Wolf03} estimated the LF for faint QSOs over the range $1.2<z<4.8$; and
\citet{Jiang06} estimated the QSO LF over the range $0.5<z<3.6$ by using a deep
survey of faint QSOs in the SDSS.  Obscured (or type 2) QSOs may be missed
in the optical surveys but can be detected in hard X-ray surveys.
\citet{LaFranca05} use 508 AGNs to estimate the hard X-ray LF (HXLF;
$2-10$~keV) over the range $0<z<2.5$ by combining data from XMM-Newton (Lockman
hole) and the Chandra Deep Field (CDF).  \citet{Barger05} use a
spectroscopically complete deep and wide-area Chandra survey to estimate the
HXLF ($2-8$~keV) over the range $0<z<5$.  \citet{Silverman08} measure the HXLF
($2-8$~keV) up to $z\sim 5$ with fewer uncertainties by combining the
observations from the CDF and the Chandra Multiwavelength Project. Combining
all these observations, the time integrals of the QSO LF are estimated in
\S~\ref{sec:QSOLF}.

In \S~\ref{sec:models}, we assume several models for the luminosity evolution
history of individual QSOs, i.e., ${\cal L}(\tau;\mbh)$, and then apply the
models and the observational BHMF and QSO LF to equation (\ref{eq:relation}) to
give constraints on the growth of individual MBHs and the associated
parameters, specifically, the efficiency (mainly determined by
the spin of a MBH), the lifetime of nuclear activity, and
the long-term evolution of disk accretion etc.  We find that a reference model
for the luminosity evolution history of individual QSOs, i.e., an initial rapid
accretion phase with a rate close to the Eddington limit and then a following
power-law declining phase set by the self-similar long-term evolution of disk
accretion ($\dot{\bh}\propto \tau^{-\gamma}$, and $\gamma\sim 1.2-1.3$), can
satisfy the extended So{\l}tan argument (eq.~\ref{eq:relation}) well. Using the
reference model for ${\cal L}(\mbh,\tau)$, we discuss the role of obscuration
in the BH growth history in \S~\ref{sec:obscuration} and find that obscuration
is unlikely to be solely an evolutionary effect. The luminosity (or
accretion-rate) evolution constrained by the extended So{\l}tan argument also
implies a distribution of Eddington ratios (i.e., the accretion rate in units
of the Eddington limit) in QSOs. In \S~\ref{sec:accrete}, we particularly
discuss its distribution expected from the models and compare them with
observations.  In \S~\ref{sec:BHmerger}, by using toy models, we discuss the
effects of BH mergers on our results, which are shown to be insignificant. In
\S~\ref{sec:discon}, we discuss further implications of the luminosity
evolution obtained from the extended So{\l}tan argument. Together with the QSO
LF, ${\cal L}(\tau;\mbh)$ can be used to further derive the BHMF at redshift
$z$ and the triggering rate of nuclear activity ${\cal G}(z; \mbh)$.  Given ${\cal
L}(\tau;\mbh)$ and ${\cal G}(z;\mbh)$, many statistical properties of QSOs can be
inferred and comparison of them  with observations may further deepen our
understanding of the growth of MBHs.  Conclusions are given in
\S~\ref{sec:conclusion}.

In this paper we set the Hubble constant as $H_0=100h\kms\Mpc^{-1}$; and if not
otherwise specified, the cosmological model used is $(\Omega_m,
\Omega_{\Lambda},h)=(0.3,0.7,0.7)$.

\section{The mass function of MBHs at $\lowercase{z}=0$} 
\label{sec:BHMF}

Studies of central MBHs in nearby galaxies have revealed strong correlations
between the BH mass and the velocity dispersion (or luminosity, or other
properties) of the hot stellar component of the host galaxy
\citep[e.g.,][]{KR95, Magorrian98, FM00, Gebhardt00, Tremaine02, HR04, MH04,
Graham01, AR07, Hopkins07b}. We first present several latest fits of these
correlations (i.e., the $\mbh-\sigma$ relation and the $\mbh-L\bulge$ relation)
and then present the observational velocity-dispersion (and luminosity)
distribution of nearby galaxies. By combining them, we estimate the local BHMF
in \S~\ref{subsec:BHMF}.

\subsection{The $\mbh-\sigma$ and $\mbh-L\bulge$ relations}\label{subsec:correlation}

\citet{Lauer07a} show that the logarithm of the BH mass at a given velocity
dispersion $\sigma$ has a mean value given by 
\begin{eqnarray}
\langle\log\mbh|\log\sigma\rangle&=&(8.29\pm0.07)+(4.13\pm 0.32)\times 
\nonumber \\
& & \log\left(\frac{\sigma}{200\kms}\right),
\label{Lauer1}
\end{eqnarray}
which is fitted in the $(\log\mbh,\log\sigma)$ space. The mean value at
a given $V$-band absolute magnitude $M_V$ is given in the same paper as
\be
\langle\log\mbh|M_V\rangle=(8.67\pm0.09)-\frac{(1.32\pm0.14)}{2.5}(M_V+22).
\label{Lauer}
\ee
The intrinsic scatters around the relations above are not reported in
\citet{Lauer07a}. (Hereafter the intrinsic scatters in $\log\mbh$ are noted as
$\Delta_{\mbh-\sigma}$ and $\Delta_{\mbh-L_{\rm bulge}}$ for the $\mbh-\sigma$
relation and $\mbh-L_{\rm bulge}$ relation, respectively.)  Based on the same
sample, \citet{Tremaine02} estimate that the intrinsic scatter in $\log\mbh$
for the $\mbh-\sigma$ relation, i.e., $\Delta_{\mbh-\sigma}$, should be not
larger than $0.25-0.3$~dex. The latest fit of the $\bh-\sigma$ relation by
\citet{Hu08}, which is consistent with that given by \citet{Lauer07a} on the
zero point and the slope, also gives an upper limit to the intrinsic scatter
$\sim 0.25$~dex. Note also the zero point in equation (\ref{Lauer1}) is larger
than that obtained by \citet{Tremaine02} by $0.10$~dex, but roughly consistent
with statistical errors. 

The estimates of the $\mbh-\sigma$ and $\mbh-L\bulge$ relations in
\citet{Bernardi07} are given by:
\begin{eqnarray}
\langle\log\mbh|\log\sigma\rangle&=&(8.21\pm0.05)+(3.83\pm 0.10)\times \nonumber \\
& & \log \left(\frac{\sigma}{200\kms}\right),
\label{Bernardi1}
\end{eqnarray}
and
\be
\langle\log\mbh|M_r\rangle=(8.57\pm0.10)-\frac{(1.30\pm0.10)}{2.5}(M_r+22), 
\label{Bernardi}
\ee
with intrinsic scatters not larger than $0.22 \pm 0.05$~dex and $0.33 \pm
0.08$~dex, respectively.  Another set of fits to equation (\ref{Bernardi}) by
the same authors \citep{Tundo07} finds a slope $1.30\pm0.15$ and the zero point
is $8.68\pm0.10$, consistent with statistical errors. If we convert $M_r$ to
$M_V$ with $M_r=M_V-0.37$ adopted for early-type galaxies \citep{Fukugita96},
we find the zero point in equation (\ref{Bernardi}) is larger than that in
equation (\ref{Lauer}) by $0.09$~dex.

The typical difference in the zero point among different sets of fits to the
$\mbh-\sigma$ (or $\mbh-L_{\rm bulge}$) relation is $\la 0.10$~dex, which is
roughly consistent with the statistical errors in the zero point estimation.
The difference in the slope among different sets of fits to the $\mbh-\sigma$
relation is quite large compared to the statistical errors in the fits, for
example, it is $4.13\pm 0.32$ in \citet{Lauer07a}, $3.83\pm 0.10$ in
\citet{Bernardi07}, and $4.86\pm 0.43$ in \citet{FF05} (for details of the
difference in the slope see discussions in \citealt{Tremaine02}). Note that
\citet{AR07} investigate an alternative relation to the $\mbh-\sigma$ relation;
and they find that the relation between the MBH mass and the bulge
gravitational binding energy is as good as the $\mbh-\sigma$ relation in
predicting MBH mass but with a slope much more stable regarding of changes in
the fitting algorithm. A detailed study by \citet{Novak06} demonstrates that
the upper limit to the intrinsic scatter is $\sim 0.2-0.3$~dex in the
$\mbh-\sigma$ relation and is $\sim 0.3-0.4$~dex in the $\mbh-L\bulge$ relation
for currently available samples.  Below we adopt $\Delta_{\mbh-\sigma}\sim
0.3$~dex relation and $\Delta_{\mbh-L_{\rm bulge}}\sim 0.4$~dex if not
otherwise specified. 

Among the subtle differences in zero points, slopes and intrinsic scatters of
those relations estimated by different groups, the intrinsic scatter would be
the most significant one for the purpose of studying MBH growth, because it may
affect the estimates of the abundance of MBHs at the high-mass end ($\ga
10^9\msun$) by orders of magnitude (as shown in Fig.~\ref{fig:lmf} below; see
also discussions in \citealt{YL04a, Marconi04, Lauer07a, Tundo07}), and
this abundance is crucial for our understanding of the growth of the most
massive BHs in bright QSOs.  

\subsection{The velocity-dispersion distribution function and 
the luminosity function of nearby galaxies} \label{subsec:sigma}

We define $n_\sigma(\sigma,t)$ as the comoving velocity-dispersion function of
the hot stellar components of local galaxies so that
$n_\sigma(\sigma,t_0)\d\sigma$ represents the number density of local galaxies
in the range $\sigma\rightarrow\sigma+\d\sigma$.  The velocity-dispersion
distribution $n_\sigma(\sigma,t_0)$ includes the contribution from both
early-type galaxies $n\early_\sigma(\sigma,t_0)$ and bulges of late-type
galaxies $n\late_\sigma(\sigma,t_0)$, that is,
\be 
n_\sigma(\sigma,t_0)=n\early_\sigma(\sigma,t_0)+n\late_\sigma(\sigma,t_0).
\label{eq:veldisp}
\ee

\begin{itemize}
\item
The velocity-dispersion distribution in early-type galaxies has been estimated
by recent studies of a sample of early-type galaxies at $z<0.3$ obtained by the
SDSS (see eq.~4 in \citealt{Sheth03}, and \citealt{Bernardi03}):
\noindent
\be
n\early_\sigma(\sigma,t_0)=\phi_*\left(\frac{\sigma}{\sigma_*}\right)^{\alpha}
\frac{\exp\left[-(\sigma/\sigma_*)^{\beta}\right]}{\Gamma(\alpha/\beta)}
\frac{\beta}{\sigma},
\label{eq:earlydisp}
\ee
where the best-fit values of $(\phi_*,\sigma_*,\alpha,\beta)$ are $(0.0020
\pm0.0001,88.8\pm17.7,6.5\pm1.0,1.93\pm0.22)$, $\phi_*$ is the comoving number
density of local early-type galaxies in units of $h^3_{0.7}\Mpc^{-3}$, and
$\sigma_*$ is in units of $\kms$. The brightest cluster galaxies (BCGs) are
probably under-represented in the above sample \citep{Lauer07a}.  We correct
this by adding the number density of BCGs to equation (\ref{eq:earlydisp}),
where the number density of BCGs with $\sigma> 350\kms$ is estimated from the
sample of \citet{Bernardi06} as done in \citet{Lauer07a}.  If the scatter in
the $\mbh-\sigma$ (or $\mbh-L_{\rm bulge}$) relation is not significantly
smaller than $0.3$~dex (or $0.4$~dex), this correction is not significant
because most of high-mass MBHs (larger than a few $10^9\msun$) actually come
from `modest' galaxies with unusually large MBHs for their velocity dispersions
or luminosities (see the dependence of the BHMF on different values of the
scatter in Fig.~\ref{fig:lmf}; see also \citealt{Lauer07b}).

\item 
The velocity-dispersion distribution in late-type galaxies
$n_\sigma\late(\sigma,t_0)$ may be estimated in the following ways. (i)
Following \citet{Sheth03}, the LF of the late-type galaxies can be obtained by
subtracting the LF of the early-type galaxies \citep{Bernardi03} from the total
LF of all galaxies \citep{Blanton03}.  (ii) Following \citet{Sheth03}, the
distribution of the circular velocity $v\c$ in late-type galaxies may be
obtained by using the LF of the late-type galaxies obtained above and the
following Tully-Fisher relation \citep{Giovanelli}
\be
\log\left(\frac{2v\c}{\kms}\right)=1.00-(M_I-5\log h_{0.7})/7.95,
\label{eq:tullyfisher}
\ee
where $M_I$ is the absolute magnitude of the galaxies in the $I$ band, with
accounting for the intrinsic scatter around relation (\ref{eq:tullyfisher}) and
the inclination effects of galaxies (see details in \citealt{Sheth03}).  (iii)
The velocity-dispersion function of late-type galaxies can be obtained by using
the circular-velocity distribution of the late-type galaxies obtained above and
the following relation between the circular velocity and the velocity
dispersion of the bulge component (see eq.~3 in \citealt{Baes03}, and also
\citealt{Ferrarese02}):
\begin{eqnarray}
\log\left(\frac{v\c}{200\kms}\right)&=&
(0.96\pm0.11)\log\left(\frac{\sigma}{200\kms}\right)\nonumber \\
& & +(0.21\pm0.023).
\label{eq:vcircsigma}
\end{eqnarray}
The intrinsic scatter of relation (\ref{eq:vcircsigma}) is small ($<0.15\dex$,
see Fig.~1 in \citealt{Baes03}) and will be ignored in our calculations. We
could also simply use $\sigma=v\c/\sqrt{3}$ (e.g., see problem~4.35 in
\citealt{BT08}) to estimate $\sigma$, which only induces a slight difference in
estimating the BHMF.  Relation (\ref{eq:vcircsigma}) may not hold for
$\sigma<80\kms$, which corresponds to $\mbh \la 4\times 10^6\msun$ according to
the $\mbh-\sigma$ relation above (eqs.~\ref{Lauer1} and \ref{Bernardi1}), but
this is beyond the main range which we focus on in \S~\ref{sec:models}.  Note
that the local BHMF for mass $\mbh \ga 4\times 10^7\msun$ is dominated by the
early-type galaxies (see also Fig.~\ref{fig:lmf} in \citealt{YL04a}).  

\end{itemize}

The LF of galaxies is conventionally described by the \citet{Schechter76}
function: \begin{eqnarray} \Phi(M)&=&0.4\ln(10){\phi}_{\ast}
10^{-0.4(M-M_*)(\alpha+1)}\times \nonumber\\ & & \exp[-10^{-0.4(M-M_*)}],
\end{eqnarray} where $\Phi(M)dM$ gives the comoving number density of galaxies
with absolute magnitude in the range $M\rightarrow M+dM$.  Based on
observations by the SDSS \citep{Blanton03}, the best fit parameters
[${\phi}_*/(10^{-2}h^3_{0.7}\Mpc^{-3})$, $M_*-5\log h_{0.7}$, $\alpha$] of the
LFs are ($6.36\pm0.23$, $-18.62\pm0.02$,$-0.89\pm0.03$) in the $g$ band and
($4.34\pm0.12$,$-19.67\pm0.01$,$-1.05\pm0.01$) in the $r$ band, respectively.
Here $M$ is the absolute magnitude of a galaxy (not just of its hot stellar
component). We can crudely estimate the luminosity of the hot stellar component
of a galaxy, for which the relations in equations (\ref{Lauer}) and
(\ref{Bernardi}) are applied, from the total luminosity of the galaxy $L\gal$
by setting $L_{\rm bulge}=(L\gal/L_*)/(1+L\gal/L_*)L\gal$
\citep[e.g.,][]{Tundo07}. With this modification, the BHMF can be estimated
using either the $\mbh-M_V$ relation (eq.~\ref{Lauer}) or the
$\mbh-M_r$ relation (eq.~\ref{Bernardi}) and the galaxy LF in the $g$
band (with a color correction of $g=M_V+0.41$; \citealt{Fukugita95}) or
the $r$ band.

\subsection{$n_{\bh}(\mbh,t_0)$}\label{subsec:BHMF}

\begin{figure*}
\begin{center}
\includegraphics[width=0.9\textwidth,angle=0]{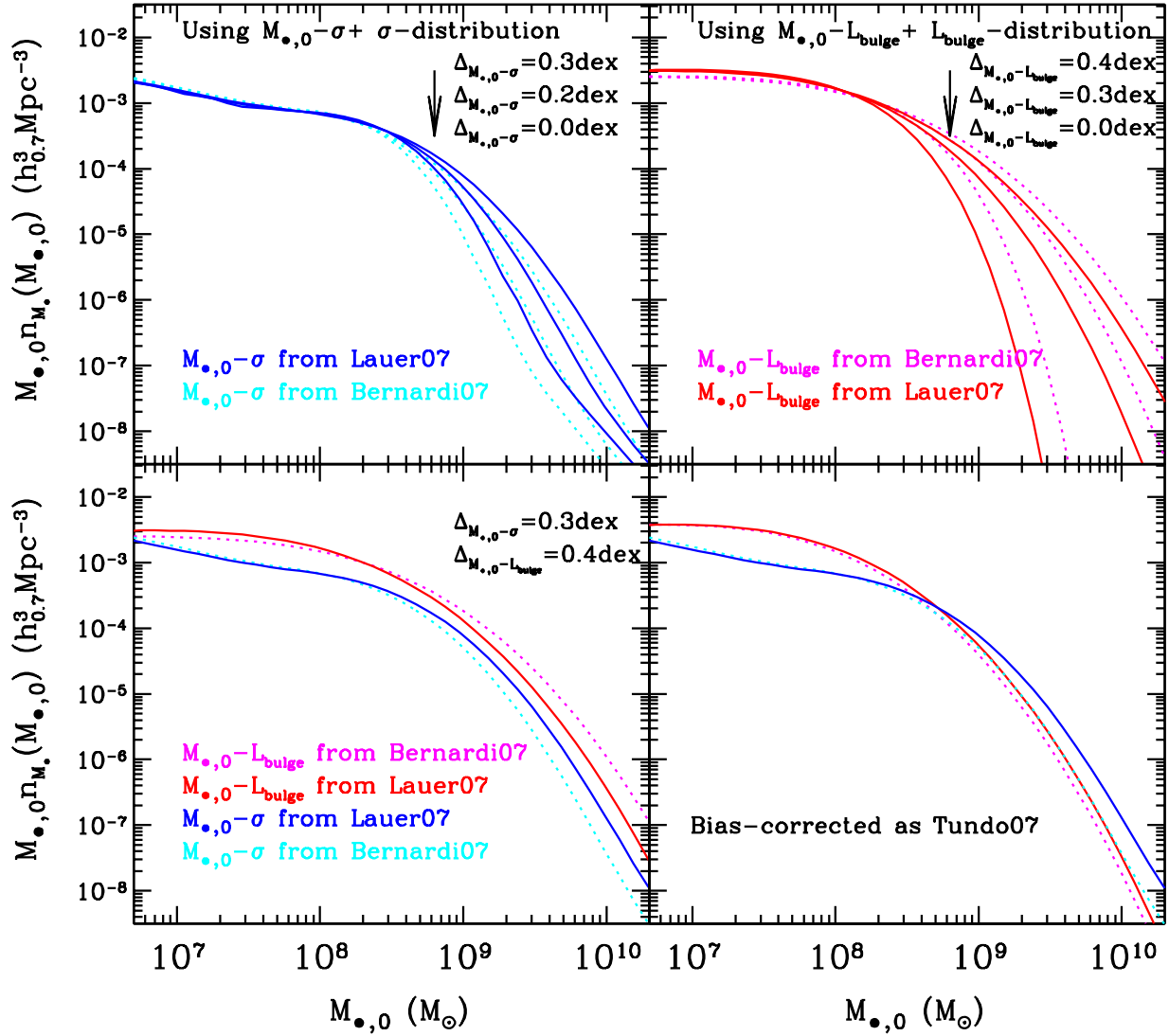}
\caption{ 
The local BHMF obtained from the velocity-dispersion/luminosity distribution
function of nearby galaxies and the MBH mass versus
velocity-dispersion/luminosity relation (see details in \S~\ref{sec:BHMF}).
Upper left panel: the solid lines represent $\mbh n_{\bh}(\mbh,t_0)$ obtained
from the $\mbh-\sigma$ relation given in \citet[][here Lauer07]{Lauer07a}, with
assumed intrinsic scatters $\Delta_{\bh-\sigma}=0.3$, $0.2$, and $0$~dex from
top to bottom, respectively; while the dotted lines represent the $\mbh
n_{\bh}(\mbh,t_0)$ obtained from the $\mbh-\sigma$ relation given in
\citet[][here Bernardi07]{Bernardi07} with the same assumed intrinsic scatters.
The velocity-dispersion distribution is obtained from equation
(\ref{eq:veldisp}).  Upper right panel: BHMF estimated from the galaxy
luminosity function and the $\mbh-L\bulge$ relation with assumed intrinsic
scatter $\Delta_{\bh-L_{\rm bulge}}=0.4$, $0.3$ and $0$~dex from top to bottom,
respectively.  The solid lines represent the BHMF obtained from the
$\mbh-L_{\rm bulge}$ relation given in Lauer07, while dotted lines for the
relation given in Bernardi07.  The bulge luminosity of a galaxy is used here
by modifying the galaxy luminosity function to the bulge luminosity
function as shown in \S~\ref{subsec:sigma}.  Bottom left panel: comparison of
the BHMF obtained from the $\mbh-\sigma$ relation and that obtained from the
$\mbh-L_{\rm bulge}$ relation for the relations estimated either in Lauer07a or
in Bernardi07.  The intrinsic scatter for the $\mbh-\sigma$ (or $\mbh-L_{\rm
bulge}$) relation is assumed to be $\Delta_{\bh-\sigma}=0.3$~dex (or
$\Delta_{\bh-L_{\rm bulge}}=0.4$~dex), which is taken as the most probable
number in this paper.  Bottom right panel: similar to the bottom left panel but
with corrections for bias as suggested by \citet[][here Tundo07]{Tundo07}.  
} 
\label{fig:lmf} 
\end{center}
\end{figure*}

We show in Figure \ref{fig:lmf} the BHMF obtained by combining the
$\mbh-\sigma$ (or $\mbh-L\bulge$) relation with the velocity-dispersion (or
luminosity) distribution function of local galaxies (e.g., see eq.~44 in
\citealt{YL04a}). Our calculations show that the uncertainties in the intrinsic
scatter of the $\mbh-\sigma$ (or $\mbh-L_{\rm bulge}$) relation may affect
estimates of the BHMF significantly at the high-mass end (see
Fig.~\ref{fig:lmf}).  To illustrate this effect, we assume that the intrinsic
scatters in the $\mbh-\sigma$ (or $\mbh-L_{\rm bulge}$) relation by
\citet{Lauer07a} (eqs.~\ref{Lauer1} and \ref{Lauer}) and by \citet{Bernardi07}
(eqs.~\ref{Bernardi1} and \ref{Bernardi}) are 0, 0.2 and 0.3~dex (or 0, 0.3 and
0.4~dex), respectively.  With the intrinsic scatter of the $\mbh-\sigma$ (or
$\mbh-L_{\rm bulge}$) relation $\sim 0.3$~dex (or $\sim 0.4$~dex), the
estimated abundance of MBHs at the high-mass end ($\ga 10^9\msun$) is larger
than that estimated from a zero intrinsic scatter by orders of magnitude (see
the upper panels of Fig.~\ref{fig:lmf}). The difference in the slope and the
zero point among different sets of fits to the $\mbh-\sigma$ (or $\mbh-L_{\rm
bulge}$) relation may also affect the estimates of the abundance of MBHs at the
high-mass end, but its effects are substantially less significant compared to
that of the intrinsic scatter (see Fig.~\ref{fig:lmf} and also
\citealt{YL04a}).  

As shown in the bottom left panel of Figure~\ref{fig:lmf}, the abundance of
MBHs estimated from the $\mbh-L\bulge$ relation is larger than that from the
$\mbh-\sigma$ relation roughly by a factor $\sim 2$ if both relations are
adopted from \citet{Lauer07a} (see also discussions in \citealt{Lauer07a} and
\citealt{Tundo07}), but the shapes are similar. One possible reason for this
discrepancy in abundance is that the local MBH sample used to derive the
$\mbh-L\bulge$ relation is biased relative to the SDSS galaxy sample as
discussed in \citet{YT02} and \citet{Bernardi07}.  (The other possibility is
systematic differences in measurements of luminosity or velocity dispersion
between other surveys and the SDSS.)  If we correct this `bias' with the recipe
introduced in \citet{Tundo07}, then the BHMF estimated from the $\mbh-L\bulge$
relation is almost the same as that estimated from the $\mbh-\sigma$ relation
at the high-mass end ($\mbh\ga$ a few $10^8\msun$), as shown in the bottom
right panel of Figure~\ref{fig:lmf}. The remaining discrepancy at the low-mass
end is possibly due to uncertainties in the estimation of the bulge luminosity
from the total luminosity for late-type galaxies. For example, recent studies
by \citet{Laurikainen05} and \citet{GW08} have shown that the bulge-to-total
luminosity ratio (B/T ratio) is around 0.24 for S0 galaxies, which is
substantially smaller than the previous estimates ($\sim 0.6$; e.g.,
\citealt{Fukugita98}). According to these new estimates, the B/T ratio adopted
in \citet{Tundo07} may be an overestimate at least for S0 galaxies, and thus
the BHMF at the low-mass end $\la 10^8\msun$ is probably substantially
overestimated. (The B/T ratio adopted in other estimates of the BHMF may be
also overestimated; e.g., \citealt{Marconi04}.) It is anticipated that the
BHMF at the low-mass end estimated by using the $\mbh-L_{\rm bulge}$ relation
will be closer to that estimated by using the $\mbh-\sigma$ relation if
adopting a more realistic B/T ratio for spiral galaxies.  In
\S~\ref{sec:models}, we adopt the BHMF obtained from the $\mbh-\sigma$ relation
given by \citet{Lauer07a} with an intrinsic scatter of $0.3$~dex as the
reference BHMF, if not otherwise specified.

In addition to the uncertainty on the local BHMF due to the intrinsic scatter
in the $\mbh-\sigma$ (or $\mbh-L_{\rm bulge}$) relation, the local BHMF suffers
other uncertainties, in particular, the uncertainties in estimating the
$\mbh-\sigma$ (or $\mbh-L_{\rm bulge}$) relation [e.g., due to (1) limited mass
range and small samples; (2) being restricted to ellipticals, and little is
known about late-type galaxies; (3) determining $\mbh$ is difficult and may be
underestimated, especially for BCGs] and the uncertainties in estimating the
velocity-dispersion (or bulge luminosity) distribution in late-type galaxies.

The total mass density of local MBHs can be estimated from the BHMF. The
differences in the zero point, the slope and the intrinsic scatter among the
relations estimated by different groups could cause at most a 20-30\%
difference in the total mass density of local MBHs (as shown in
\S~\ref{subsec:correlation}). For example, adopting the $\mbh-\sigma$ relation
given by \citet{Lauer07a} yields a total mass density of MBHs $\simeq
3.8^{+0.7}_{-0.6}\times 10^5 h_{0.7}^{2}\msun{\rm Mpc}^{-3}$, which is larger
than that obtained by \citet{YT02} by a factor of $\sim 1.3$ mainly due to the
larger zero point of the $\mbh-\sigma$ relation in \citet{Lauer07a} adopted
here. We show in Table~\ref{tab:1} a few estimates of the total mass density of
local MBHs obtained from the $\mbh-\sigma$ (or $\mbh-L_{\rm bulge}$) relation
given by different authors. The errors are obtained by accounting for the
uncertainties in the $\mbh-\sigma$ (or $\mbh-L_{\rm bulge}$) relation and the
galaxy velocity-dispersion (or the luminosity) distribution function. (For
other estimates of the total mass density of local MBHs, see Tab.~3 in
\citealt{Graham07}.) The total mass density obtained from the $\mbh-L_{\rm
bulge}$ relation is about a factor of $\sim 2$ larger than that obtained from
the $\mbh-\sigma$ relation, which is consistent with that in \citet{YT02} (see
also discussions for the reasons of this discrepancy in \citealt{Tundo07}). If
we use the recipe introduced by \citet{Tundo07} to correct the possible bias in
MBH masses estimated from the $\mbh-L_{\rm bulge}$ relation, the corrected
total mass densities are still larger than that obtained from the $\mbh-\sigma$
relation but now appears to be consistent within statistical errors (see
Tab.~\ref{tab:1}). Furthermore, considering that the B/T ratio for spiral
galaxies adopted in the estimates of total BH mass density using the
$\mbh-L_{\rm bulge}$ relation is probably an overestimate, the total
BH mass density from the $\mbh-L_{\rm bulge}$ and the galaxy LF may be actually
not much different from that estimated from the $\mbh-\sigma$ relation and the
galaxy velocity dispersion distribution function.

\begin{table}
   \begin{center}
      \begin{small}
         \caption{The total mass density of massive black holes
                  }
         \begin{tabular}{llcc}
         \hline
    Method & Reference &  Note      & $\frac{\rho_{\bullet,0}}{10^5\msun}$ \\
    \hline
    $\mbh-\sigma$        & Lauer07a   & ......         & $3.8^{+0.7}_{-0.6}$ \\
    $\mbh-\sigma$        & Bernardi07 & ......         & $3.3^{+0.5}_{-0.4}$ \\
    $\mbh-\sigma$        & FF05       & ......         & $3.6^{+0.7}_{-0.6}$ \\
    $\mbh-L_{\rm bulge}$ & Lauer07a   & ......         & $7.6^{+2.0}_{-1.7}$ \\
    $\mbh-L_{\rm bulge}$ & Bernardi07 & ......         & $8.5^{+2.4}_{-2.0}$ \\
    $\mbh-L_{\rm bulge}$ & Lauer07a   & Bias corrected & $4.9^{+1.3}_{-1.0}$ \\
    $\mbh-L_{\rm bulge}$ & Bernardi07 & Bias corrected & $4.3^{+1.2}_{-1.0}$ \\
         \hline 
         \label{tab:1}
         \end{tabular}
      \end{small}
   \end{center}
\tablecomments{The total mass density of massive black holes estimated from
                  the $\mbh-\sigma$ (or $\mbh-L_{\rm bulge}$) relation obtained
                  by different authors. The references for the $\mbh-\sigma$
                  (or $\mbh-L_{\rm bulge}$) relation are listed in Column
                  2, and Lauer07a, Bernardi07, FF05 represent \citet{Lauer07a},
                  \citet{Bernardi07} and \citet{FF05}, respectively.}
\end{table}

\section{The QSO/AGN LF in the optical and hard X-ray bands} \label{sec:QSOLF}

\subsection{The optical QSO LF}

\begin{figure}
\begin{center}
\includegraphics[width=\hsize,angle=0]{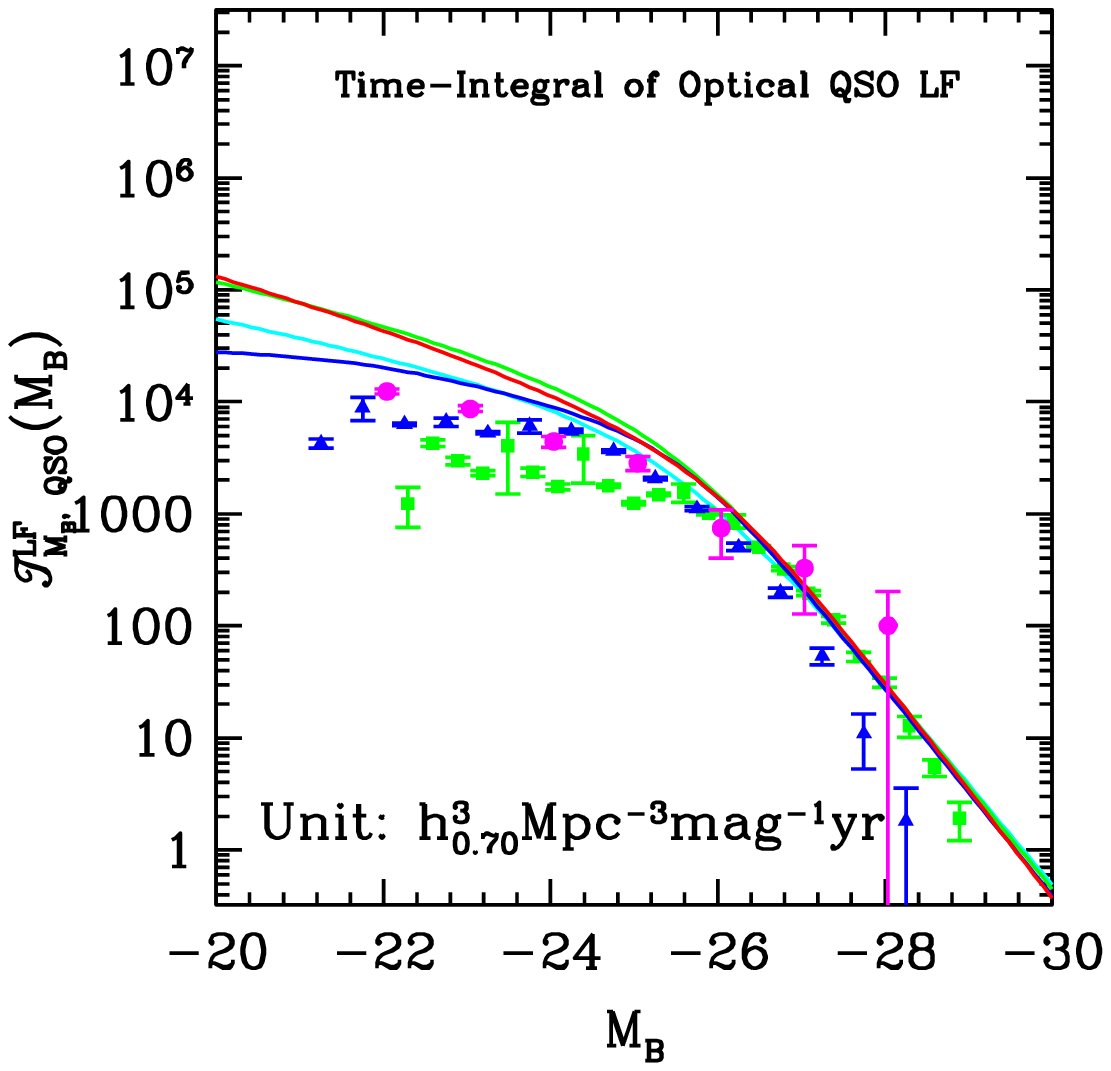}
\caption{
The time integral of the QSO luminosity function in the optical band ($B$).
The solid triangles (blue), solid circles (magenta) and solid squares (green)
represent the direct summation of the binned QSO LF (eq.~\ref{eq:sum}) obtained
from \citet{Croom04} over the redshift range $0.35<z<2.1$, \citet{Wolf03} over
the redshift range $1.2<z<4.8$ and \citet{Richards06a} over the redshift range
$0.3<z<5.0$, respectively. High-luminosity QSOs are under-represented in
\citet{Croom04} because of the redshift limit ($z<2.1$) and low-luminosity QSOs
are under-represented in the sample in \citet{Richards06a} because of
incompleteness at the faint end (especially at high redshift). The blue, cyan,
red, and green lines represent the time integrals obtained  from the fitting
formulae in \citet{Croom04}, \citet{Jiang06}, \citet{Boyle00}, and
\citet{Richards05}, respectively.  The time integrals obtained from different
fitting formulae are consistent at the high-luminosity end but show substantial
discrepancies at the low-luminosity end ($M_{\rm B}> -23$) mainly because of
the uncertainties in the faint-end slope of the QSO LF.  For example, the
faint-end slope is estimated to be $\sim 1.09$ by \citet{Croom04}, $\sim 1.25$
in \citet{Jiang06}, but $\sim 1.58$ by \citet{Boyle00} and $\sim 1.45$ by
\citet{Richards05}. See also in Fig.~\ref{fig:QSOILFadd}.
}
\label{fig:QSOILF}
\end{center}
\end{figure}

The optical QSO LF was first estimated by
\citet{Schmidt68} and \citet{SG83}, and it has been investigated extensively
since then. The shape and evolution of the QSO LF has been well, though
not perfectly, constrained due to recent surveys with unprecedentedly 
large redshift and luminosity spans \citep[e.g.,][]{Boyle00,Wolf03,
Croom04,Richards05,Richards06a,Jiang06,Fontanot07,Siana07}. Using a sample
of more than 15,000 QSOs at redshift $z<2.5$ from 2Qz and 6Qz, 
\citet{Croom04} obtained the binned QSO LF $\Psi_{M}(M_{{\rm b_J},i},z_j)$ 
over the range $0.4<z<2.1$ and the magnitude range $M_{\rm b_J}<-22.5$, 
where $M_{{\rm b_J},i}$ is the $i$th bin of the absolute magnitude and $z_j$
is the $j$th bin of the redshift. For some high-redshift bins,
the binned QSO LF at low luminosity is not available because of the flux
limit of the surveys. The time integral of the QSO LF can be estimated through
direct summation by multiplying the binned QSO LF by the cosmic time duration as
\be
{\cal T}'_{M_{\rm b_J,QSO}}=\sum_j \Psi_{M}(M_{{\rm b_J},i},z_j)\Delta t(z_j),
\label{eq:sum}
\ee
where $\Delta t(z_j)$ is the cosmic time interval corresponding to the redshift
bin $z_j$, $\Psi$ is assumed to be 0 outside observational bins, and the prime
$'$ indicates the value obtained by summation over bins---in contrast the
variable ${\cal T}$ without prime (see eq.~\ref{eq:calT}) represents the time
integral of a continuous fit to the QSO LF. These summations only give lower
limits to the time integral of the QSO LF because the binned QSO LF, especially
in the low-luminosity bins, does not extend to high enough redshift to include
all QSOs.  \citet{Richards06a} obtained the binned QSO LF over a larger
redshift range ($0.3<z<5$) using a homogeneous statistical sample of 15,343
QSOs drawn from SDSS Data Release 3. Unfortunately, the SDSS survey is shallow
so the binned QSO LF can only be determined at the bright end. As a complement
to the above estimates, the QSO LF for faint QSOs over the range $1.2<z<4.8$
was estimated by \citet{Wolf03} using the COMBO-17 data; by \citet{Jiang06}
over the range $0.5 < z <3.6$ using a deep survey of faint QSOs in the SDSS; by
\citet{Fontanot07} in the redshift range $3.5<z<5.2$ by combining the data from
the Great Observatories Origins Deep Survey (GOODS) and the SDSS; and by
\citet{Siana07} at the redshift range $2.83<z<3.44$ using the data from the
Spitzer Wide-area Infrared Extragalactic (SWIRE) Legacy Survey.  In
Figure~\ref{fig:QSOILF}, the direct summations (eq.~\ref{eq:sum}) are shown for
the binned QSO LF from \citet[][blue triangles]{Croom04}, \citet[][magenta
circles]{Wolf03} and \citet[][green squares]{Richards06a}, respectively (the
$M_i$ magnitude in \citealt{Richards06a} and $M_{145}$ magnitude in
\citealt{Wolf03} are all converted to $M_B$ magnitude by $M_{\rm B}\simeq
M_i(z=2)+0.80$ and $M_{\rm B}= M_{145}+1.75$, see \citealt{Richards06a} and
\citealt{Wolf03}). At the high-luminosity end, the estimate 
from \citet{Croom04} is substantially smaller than that from
\citet{Richards06a} which emphasizes the significance of the contribution from
high-redshift QSOs.  At the low-luminosity end, the estimates from
\citet{Richards06a} are smaller than those from others because the
\citet{Richards06a} sample is shallower and the majority of faint QSOs are not
included.

\begin{figure}
\begin{center}
\includegraphics[width=\hsize,angle=0]{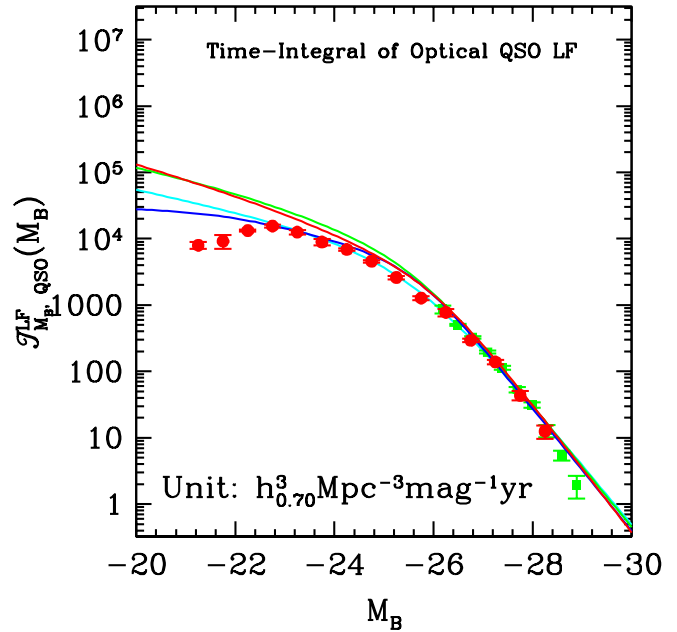}
\caption{
The time integral of the QSO LF in the optical band ($B$).  Similar to
Fig.~\ref{fig:QSOILF}, but the points are obtained by combining the binned
optical LFs given by different surveys over different redshift and luminosity
ranges.  At each redshift bin with data available, the binned optical LFs
obtained from the largest sample are adopted, and interpolations of the data
points over magnitudes at a given redshift are used. The points, which should
be lower limits to the time integrals of the QSO LF, are quite consistent with
that obtained from the fitting formulae of  the QSO LF at $M_{\rm B}<-24$
(solid lines). At the faint end, the direct summations are substantially
smaller than those estimated from the continuous fitting formulae with
extrapolations to higher redshift and lower luminosities (which may be due to
the incompleteness of the samples). The five green squares represent the
brightest QSOs obtained from \citet{Richards06a} only and they are consistent
with the trend of the red  points. The blue, cyan, red and green lines
represent the time-integrals obtained from the fitting formulae in
\citet{Croom04}, \citet{Jiang06}, \citet{Boyle00} and \citet{Richards05},
respectively.
}
\label{fig:QSOILFadd}
\end{center}
\end{figure}

We combine these binned QSO LFs obtained by different surveys over different
redshift and luminosity ranges \citep{Croom04,Wolf03,Richards06a,
Jiang06,Fontanot07,Siana07}, to cover luminosity and redshift ranges as large
as possible. The basic rule is that the binned QSO LF from the largest sample
are adopted at each redshift bin with data available and interpolations of the
data points over magnitudes at a given redshift are used.  The red points in
Figure~\ref{fig:QSOILFadd} are the estimated ${\cal T}'_{M_{\rm B,QSO}}$ with
mean magnitude corresponding to that in \citet{Croom04}. At bright magnitudes,
most of the points cover the range $0.4<z\la5$ but the two points with faintest
magnitudes only cover the range $0.4<z<1.0$.  In addition, the five green
squares represent the brightest QSOs obtained from \citet{Richards06a} only
and are consistent with the trend of the red points.

The optical QSO LF is frequently fitted with a double power law:
\be
\Psi_{M}(M,z)=\frac{\Psi_M^*}
{10^{0.4(\beta_1+1)[M-M^*(z)]}+10^{0.4(\beta_2+1)[M-M^*(z)]}},
\label{eq:QSOLF}
\ee
where $\Psi_{M}(M,z)dM$ is the comoving number density of QSOs with
absolute magnitude in the range [$M, M+dM$] at redshift $z$.  That is,
the evolution of the QSO LF can be characterized by three functions of
redshift: the slopes at both the high-luminosity ($\beta_1$) and the
low-luminosity ends ($\beta_2$) and the break luminosity (corresponding to
$M^*$).  \citet{Boyle00}, \citet{Croom04}, \citet{Richards05}, and
\citet{Jiang06} all use this functional form to fit their data sets from 2dF
and SDSS, except that \citet{Jiang06} introduced additional density evolution
to the QSO LF at high redshift ($z>2.0$). Adopting their best-fit models, the
time integral of the QSO LF, 
\be
{\cal T}_{M_{\rm QSO}}=\int \Psi_{M}(M,z) \frac{dt}{dz} dz, 
\label{eq:calT}
\ee
is obtained by integrating the QSO LFs over the range $0<z<8$. This function
is  shown in Figures \ref{fig:QSOILF} and \ref{fig:QSOILFadd}.
There are some differences in the model parameters among the best-fit models
for different samples. For example, \citet{Croom04} obtained a slope of
$\beta_2\sim 1.09$ at the faint end (blue line), but \citet{Richards05}
obtained a steeper slope ($\beta_2 \sim 1.45$; green line) using a sample from
the 2dF-SDSS LRG and QSO survey (2SLAQ) with a flux limit of one magnitude
fainter, which is roughly consistent with that obtained by \citet{Boyle00}
($\beta_2\sim 1.58$; red line).  \citet{Jiang06} also obtained a shallower
slope ($\beta_2\sim 1.25$; cyan line) with a deep survey in the SDSS, which is
similar to that ($\beta_2=1.24$) found by \citet{Hunt04} at redshift $z\sim 3$.
At high redshift ($z\ga 3$), the estimate of the faint-end slope by
\citet{Fontanot07} is consistent with $\beta_2=1.45$ but may have a high
probability to be as steep as $\beta_2=1.71$, and \citet{Siana07} obtained
$\beta_2=1.42$, which is not inconsistent with values measured at lower
redshift \citep[e.g.,][]{Richards05,Boyle00}. The differences in $\beta_2$ are
the primary reason for the differences in ${\cal T}_{M_{\rm B,QSO}}$ at the
faint end (see Figs.~\ref{fig:QSOILF} and \ref{fig:QSOILFadd}).  (Below we
choose $\beta_2\sim 1.45$ as the best estimate of the faint end of the QSO LF
in \S~\ref{sec:obscuration}.) At the high luminosity end, the direct summations
from the combination of the binned QSO LFs (according to eq.~\ref{eq:sum}),
which should be a lower limit to the time integrals, are quite consistent with
the integration obtained from extrapolations of the best-fit analytic models,
which may suggest that the estimates of ${\cal T}_{M_{\rm B,QSO}}$, at least at
the high-luminosity end, are quite secure.

\subsection{X-ray AGN LF}\label{subsec:XAGNLF}

The advantage of counting QSOs/AGNs in X-rays is that relatively low-luminosity
AGNs and obscured (type-2) AGNs, which may be missed in optical surveys, can be
unambiguously detected in deep X-ray surveys even at large redshift. Although
the number of QSOs/AGNs observed in X-rays ($\la 1000$) is still substantially
smaller than that observed in the optical band ($>10^4$), the X-ray AGN (XAGN)
LF can be estimated with considerable accuracy \citep[e.g.,][]{Ueda03,
LaFranca05,Hasinger05,Barger05,Silverman08}.  \citet{Ueda03} estimated the hard
X-ray ($2-10$~keV) LF (HXLF), which is assumed to represent the total X-ray LF
of unobscured plus Compton-thin AGNs, from a complete sample with $\sim 257$
sources observed by ASCA (but most of their sources have redshift $z<3$).
\citet{LaFranca05} estimated the HXLF using a combined sample with $508$
sources with redshift $z\la 2.5$.  With the data from Chandra deep surveys,
\citet{Barger05} extended the estimate of the HXLF ($2-8$~keV) to higher
redshift ($3\la z\leq 5$) but with large uncertainties at this redshift range.
Combining the published data from deep surveys by Chandra (i.e., CDF-North,
CDF-South) and XMM-Newton (Lockman Hole) and rare luminous sources from the
Chandra Multiwavelength Project, \citet{Silverman08} estimated the HXLF
($2-8$~keV) at redshift $3\la z\la 5$ with much smaller uncertainties. The soft
X-ray ($0.5-2$~keV) LF recently computed by \citet{Hasinger05} is assumed to
represent the unobscured type-1 AGNs. \citet{Gilli07} demonstrated that the
soft X-ray LF obtained by \citet{Hasinger05} is actually consistent with the
HXLF obtained by \citet{Ueda03} and \citet{LaFranca05} by assuming a
distribution of absorption column densities. However, the bolometric
correction (BC) for the soft X-ray band is much more uncertain than that in the
hard X-ray band, so we shall not consider the soft X-ray LF further in this
paper.  The shape and evolution of the X-ray LF in both hard X-ray and soft
X-ray bands can be described by the ``luminosity-dependent density evolution''
model \citep[e.g.,][]{Ueda03,LaFranca05,Silverman08,Miyaji00,Hasinger05}:
\be
\Psi_{\log L}(\log L_{X},z)=\frac{d\psi(<L_{ X},z)}{d\log L_{X}}=\frac{d\psi(<L_{X},z=0)}{d\log L_{X}}e(z),
\label{eq:XLF}
\ee
where $\Psi_{\log L}(\log L_{X},z)d\log L_{X}$ is the comoving number density of QSOs with
logarithm X-ray luminosity in the range [$\log L_{X}$, $\log L_{X}+d\log L_{X}$],
\be
\frac{d\psi(<L_{X},z=0)}{d\log L_{X}}=A\left[\left(\frac{L_{X}}{L_*}\right)^{\gamma_1}+
\left(\frac{L_{X}}{L_*}\right)^{\gamma_2}\right]^{-1},
\ee
\be
e(z)=\cases{(1+z)^{p_1}, & $z\le z_c$, \cr
e(z_c)[(1+z)/(1+z_c)]^{p_2}, & $z>z_c$,
}
\ee
and
\be
z_c(L_{X})=\cases{z^*_c, & $L_{X}\ge L_a$, \cr
z^*_c(L_{X}/L_a)^\alpha, & $L_{X}<L_a$.
\label{eq:zc}
}
\ee

\begin{figure}
\begin{center}
\includegraphics[width=\hsize,angle=0]{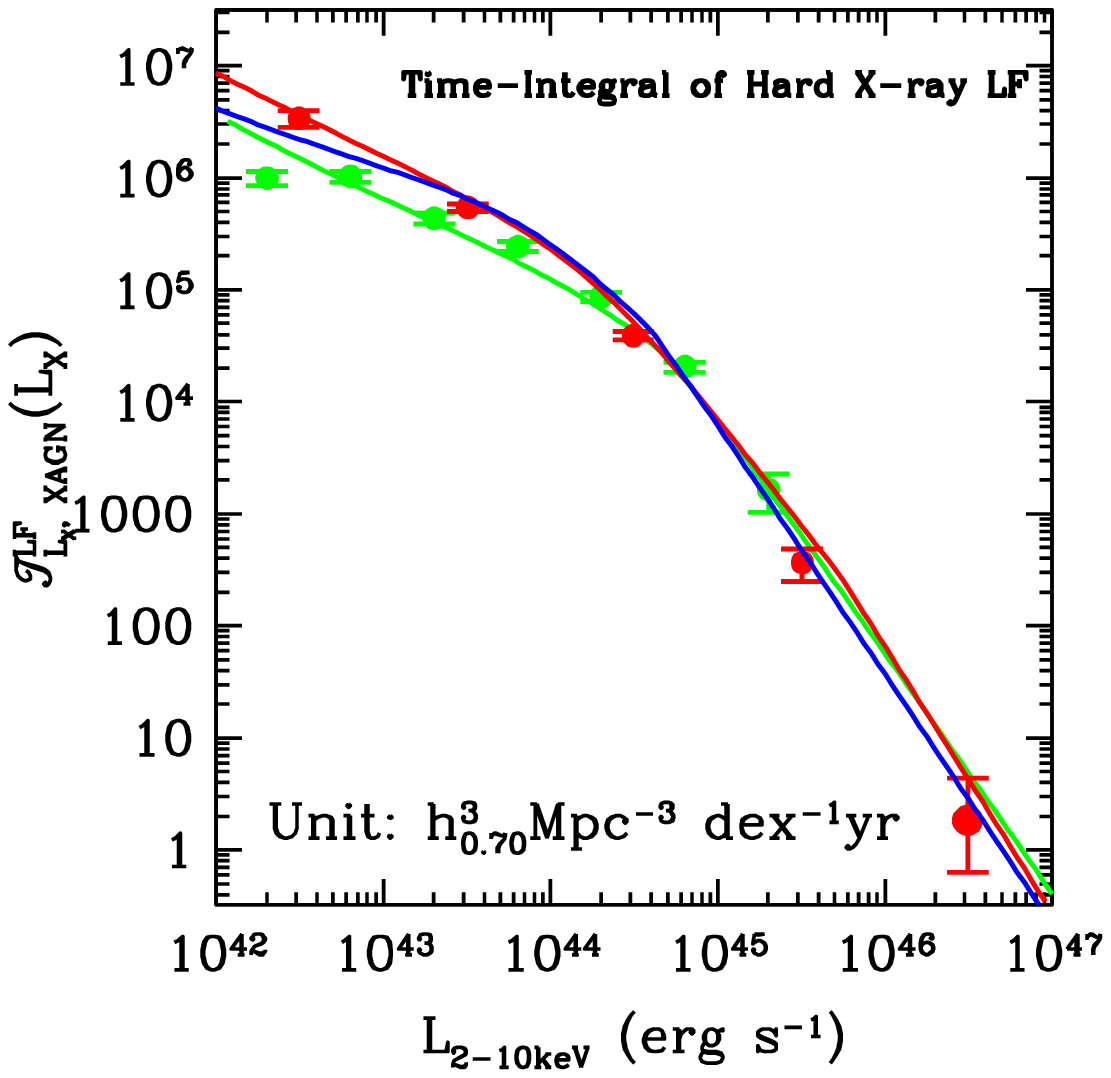}
\caption{
The time integral of the X-ray AGN LF. The red solid circles represent the
direct summations of the binned X-ray LF for the AGN sample over redshift range
$0< z< 2.5$ collected in \citet{LaFranca05}, while the green solid circles
represent the direct summations of the binned X-ray LF obtained from the AGN
sample over redshift range $0.2<z<5.5$ by surveys such as the Chandra Deep
Field (CDF) described in \citet[][see details in
\S~\ref{sec:QSOLF}]{Silverman08}.  For the low-luminosity points, the sample
may be incomplete at high redshifts and thus those points may be only lower
limits, especially for the sample in \citet{Silverman08} which may suffer from
a bias due to the optical magnitude limits in the survey. The red, blue, and
green lines represent the time integrals of the X-ray LFs in the redshift range
$0<z<8$ according to the fitting formulae obtained by \citet{LaFranca05},
\citet{Ueda03}, and \citet{Silverman08}, respectively. At the high-luminosity
end ($L_{X}>10^{44}\ergs$), the time integrals are quite consistent with
the binned data and with each other, but at the low-luminosity end ($L_{\rm
X}<10^{43}\ergs$), the one obtained from \citet{Silverman08} is smaller than
that from \citet{LaFranca05} by a factor of $\ga 2$. The estimates obtained
from \citet{LaFranca05} at the low-luminosity end ($L_{X}\la
10^{43.5}\ergs$) is taken as the best one, while at middle and
high luminosities both the estimates from \citet{LaFranca05} and
\citet{Silverman08} are taken into account.  Compton-thick sources, which are
hard to be observed even in the X-ray, are not included (for the contribution
of the Compton-thick AGNs to the time-integral of X-ray luminosity function see
discussions in \S~\ref{subsec:XAGNLF}). 
}
\label{fig:XAGNILF}
\end{center}
\end{figure}

To estimate the time integrals of the HXLF, we will use the HXLF obtained by
\citet{LaFranca05} as their AGN sample is larger than that in \citet{Ueda03}
and that obtained in \citet{Silverman08} as their X-ray LF extends to redshift
$z\sim 5$. In Figure~\ref{fig:XAGNILF}, the direct summations obtained by
multiplying the binned HXLF by the cosmic time duration in each luminosity bin
with available data in each redshift bin are shown as green and red points for
the HXLFs obtained in \citet{LaFranca05} and \citet{Silverman08}, respectively.
The $2-8$~keV luminosity in \citet{Silverman08} is converted to the $2-10$~keV
luminosity by assuming a photon index of 1.9. The time integral obtained by
integrating the HXLF over redshift $0<z<8$ (with extrapolation of the HXLF to
high redshifts and high luminosities) is shown in Figure~\ref{fig:XAGNILF} by
adopting the best-fit ``luminosity-dependent density evolution'' model of the
HXLF in \citet{Ueda03} (blue line), \citet[][model 4 in table~2]{LaFranca05}
(red line), and \citet[][model C in table~4; green line]{Silverman08},
respectively. In Figure~\ref{fig:XAGNILF}, the direct summations obtained by
multiplying the binned HXLF by the cosmic time duration, representing the
lower-limits to the time integrals of the HXLF, are quite consistent with the
time integrals obtained by integrating the best-fit X-ray LF models, which
might suggest that the majority of X-ray AGNs have been covered by current
observations although the HXLF from \citet{LaFranca05} does not cover redshift
$z>2.5$ and the sample of \citet{Silverman08} lacks high-luminosity AGNs.  At
the low-luminosity end, the time integrals obtained from the
\citet{Silverman08} HXLF is smaller than that from \citet{LaFranca05} by a
factor of $\sim 2$, which may be due to the selection bias of the magnitude
limits in the survey of the \citet{Silverman08} sample. Hereafter we take the
estimates obtained from \citet{LaFranca05} at the low-luminosity end ($L_{X}\la
10^{43.5}\ergs$) as the best estimates, while at middle and high luminosities
both the estimates from \citet{LaFranca05} and \citet{Silverman08} are taken
into account.

The X-ray cosmic background at a few to 100 keV is believed to be produced by
the integrated emission from AGNs \citep[e.g.,][]{Comastri95}. Using the
synthesis model to reproduce the observed X-ray background, a population of
Compton-thick AGNs is required to match the high energy (at $\sim 30-40$~keV)
X-ray background spectrum as measured by HEAO-1 \citep[e.g.,][]{Gilli07}. The
number density of these Compton-thick AGNs is estimated to be at most $\sim
30\%$ of the total population at $L_{X}\ga 10^{43.5}\ergs$ and not larger
than $45\%$ at lower luminosity \citep[e.g.,][]{Gilli07,MH07}. A low-limit of
the fraction of Compton-thick AGNs to the total population is probably $\sim
10-15\%$, which is set by the current observations by INTEGRAL and Swift for
bright AGNs \citep{Markwardt, Beckmann}; and locally the fraction of
Compton-thick AGN is found to be less than $20\%$ by \citet{Sazonov}.  Current
observations of the Compton-thick AGN population are insufficient to give its
(luminosity) distribution function.  We will discuss the contribution of this
population to the time-integral of AGN LF and its effect on model parameter,
but do not go into details of the Compton-thick population in the models in
\S~\ref{sec:models}.

\subsection{The BC in the optical and hard X-ray bands} 
\label{subsec:Bolcor}

The BC of a QSO is usually defined by $C_{\nu}\equiv L\bol/ (\nu L_{\nu})$,
where $\nu L_{\nu}$ is the energy radiated at the central frequency $\nu$ of a
specific band.  Based on observations from optical to hard X-rays,
\citet{Elvis94} constructed the spectral energy distributions (SEDs) for
several tens of QSOs and estimated the BC in the $B$ band, which is about
$11.8\pm 4.3$. Considering that the infrared bump in the Elvis et al.'s SED
templates was probably due to reprocessing of UV to X-ray photons by the dusty
torus rather than the intrinsic emission from the central nuclei,
\citet{Marconi04} obtained that the BC at the $B$ band is $7.9\pm 2.9$. Based
mainly on an anti-correlation between the optical-to-X-ray spectral index
($\alpha_{ox}$) and the 2500\AA ~luminosity \citep[e.g.,][]{Vignali03,
Strateva05, Steffen06}, \citet{Marconi04} and \citet{Hopkins07} re-calibrated
the SED and argued that the BC is luminosity-dependent. The BCs were derived by
\citet{Marconi04} as
\be
\log[L\bol/L_{2-10~{\rm keV}}]=1.54+0.24{\sf L}+0.012{\sf L}^2-0.0015{\sf L}^3,
\ee 
\be
\log[L\bol/\nu_{\rm B}L_{\nu_{\rm B}}]=0.80-0.067{\sf L}+0.017{\sf L}^2-0.0023{\sf L}^3,
\ee
where ${\sf L}\equiv\log L\bol-12$ and $L\bol$ is the bolometric luminosity
in units of $L_{\sun}$. \citet{Hopkins07} found
\be
\frac{L\bol}{L_{\rm band}}=c_1\left(\frac{L\bol}{10^{10}L_{\sun}}\right)^{k_1}+
c_2\left(\frac{L\bol}{10^{10}L_{\sun}}\right)^{k_2},
\label{eq:HopkinsBC}
\ee
with ($c_1,k_1,c_2,k_2$) given by (6.25, -0.37, 9.00, -0.012) for $L_{\rm band}
=L_{\rm B}$ and (10.83, 0.28, 6.08, -0.020) for $L_{\rm band}=L_{2-10~{\rm keV}}$. 
The scatter in BCs given by equation (\ref{eq:HopkinsBC}) is
\be
\sigma_{\log(L\bol/L_{\rm band})}=\sigma_1(L\bol/10^9L_{\sun})^{\beta}+\sigma_2,
\label{eq:HopkinsBCscat}
\ee
where ($\sigma_1$, $\beta$, $\sigma_2$)=($0.08$, $-0.25$, $0.06$) in the $B$
band and ($0.06$, $0.10$, $0.08$) in the hard X-ray.  The BC in hard X-ray given
by \citet{Hopkins07} is $30\%$ larger than that given by \citet{Marconi04}, and
the BC in the $B$ band given by \citet{Marconi04} is smaller than that given by
\citet{Hopkins07} by a factor of $1.5$ (or $1.8$) at $L\bol=10^{10}L_{\sun}$ (or
$L\bol=10^{14}L_{\sun}$). In this paper, we adopt the BCs for the X-ray and $B$
bands and associated scatters obtained by \citet{Hopkins07}. If the BCs given
by \citet{Marconi04} were adopted, the efficiency $\epsilon$ should be
systematically smaller than that obtained below in \S~\ref{sec:models} by a
factor of $\sim 1.3$ in order to match the time-integral of QSO/AGN LF obtained
from observations with that inferred from the local BHMF.

We note that \citet{VF07} recently investigated the SEDs of 54 AGN and found
significant spreads in the BCs. Their results suggest a relationship between
BCs in the X-ray band and Eddington ratios (see definition in
\S~\ref{sec:models}) in AGNs, with a transition at an Eddington ratio of $\sim
0.1$, below which the BC is typically $15-25$ for the $2-10$~keV luminosities
and above which the BC is typically $40-70$. Their estimates of the BC for the
optical band is approximately independent of Eddington ratio and roughly
consistent with that obtained by \citet{Hopkins07}.  We also note that simple
theoretical expectations of the BCs would be that it is not only the functions
of Eddington ratios but also the functions of MBH masses because the SED of the
disk emission depends on the MBH mass and Eddington ratio.  In addition, the
QSO/AGN variability in the hard X-ray is substantial while it is not
significant in the optical band. The X-ray variability, typically a factor of
$\sim 1.5$, introduces an additional scatter of $\sim 0.13$~dex to the BC for
the hard X-ray band \citep[see Tab.~2 in][]{VF07}.  Since a quantitative
relation between the BCs and the Eddington ratio is still premature, we shall
not consider the BCs as functions of the Eddington ratio in this work but
simply adopt equations (\ref{eq:HopkinsBC}) and (\ref{eq:HopkinsBCscat}) and
include an additional scatter due to the X-ray variability.

The time integral of the QSO LF at any given wave-band $Y$ can also be inferred
from the local BHMF as follows, provided that the BC at this band is known,
\begin{eqnarray}
{\cal T}^{\rm mod}_{Y,XAGN}&=&\int^{\infty}_{0}dL\bol\int^{\infty}_{0} 
n_{\bh}(\mbh,t_0)\tau\life(\mbh) \nonumber \\
& & P(L\bol|\mbh) P(L_Y|L\bol) d\mbh,
\label{eq:intYband}
\end{eqnarray}
where $L_{Y}$ is the luminosity at the Y-band, and $P(L_{Y}|L\bol)$ is the
probability distribution of Y-band luminosity for QSOs/AGNs with bolometric
luminosity $L\bol$ and is determined by the BCs and their scatters.

\section{Simple models for the luminosity evolution of 
individual QSOs}\label{sec:models}

In this section, we introduce three simple models for the luminosity/accretion
rate evolution of individual QSOs. These models are assumed to represent the
luminosity/accretion rate evolution averaged over an intermediate timescale
substantially smaller than the lifetime of individual QSOs, but much longer
than certain details of the evolution such as the short time variation, etc.
The parameters involved in these models will then be constrained by
observations of the local BHMF and the QSO/AGN LF through the extended
So{\l}tan argument (eq.~\ref{eq:relation}).  Because X-ray surveys are more
complete than optical surveys in the sense that obscured AGN can be detected in
X-ray surveys, we will compare the time integrals obtained from the X-ray LF
with that inferred from the local BHMF in this section, and then use the time
integral of the optical QSO LF to give constraints on obscured AGN fraction in
the optical band in \S~\ref{sec:obscuration}.

\subsection{Several fiducial parameters}\label{subsec:parameters}

We first summarize several fiducial parameters involved in the models below.

\begin{itemize}
\item The ``Eddington luminosity'' is a characteristic luminosity at which
radiation pressure on free electrons balances gravity:
\begin{eqnarray}
L\Edd(\bh)&=&\frac{4\pi G\bh m_{\rm p}c}{\sigma_{\rm T}}\nonumber \\
& \simeq & 1.26\times 
10^{46} \left(\frac{\bh}{10^8\msun}\right){\rm erg~s^{-1}},
\label{eq:Ledd}
\end{eqnarray}
where $G$ is the gravitational constant, $m_{\rm p}$ is the mass of a proton,
and $\sigma_{\rm T}$ is the cross-section of 
Thompson scattering. The Eddington luminosity is frequently assumed to be
the maximum luminosity of any object of mass $\bh$.

\item Corresponding to the Eddington luminosity, the ``Eddington accretion
rate'' is defined by: 
\begin{eqnarray} 
\dot{M}^\infty_{\rm acc,Edd}&\equiv&\frac{L\Edd}{\epsilon c^2}\nonumber \\
&=&2.22
\left(\frac{0.1}{\epsilon}\right)\left(\frac{\bh}{10^8\msun} \right)
\msun\yr^{-1}, 
\label{eq:mdotinf}
\end{eqnarray}
where $\epsilon$ is the mass-to-energy conversion efficiency; and the Eddington
growth rate of a MBH is 
\be 
\dot{M}_{\bullet,\rm Edd}=(1-\epsilon) \dot{M}^\infty_{\rm acc,Edd}.
\label{eq:mdot} 
\ee 
The efficiency $\epsilon$ is predicted to be in the range $\sim 0.04-0.31$ in
the thin disk accretion models, depending on the spin of the MBH [$\epsilon=
0.057$ for a Schwarzschild BH, and $ 0.31$ ($0.04$) for a prograde (retrograde)
rotating accretion disk around a Kerr BH with the dimensionless spin parameter
$a\sim 0.998$, the upper
limit of BH spin if the BH is spun up by accretion; \citealt{Thorne74}].
Currently, the spin of MBHs is difficult to measure directly. Theoretical
studies of the spin evolution of MBHs show that MBH spin may reach an
equilibrium point for most of its lifetime considering both accretion and
merger processes \citep[e.g.,][]{Lu96, Gammie04, Shapiro05, Volonteri05, HBK07, Nobleetal,
HB03}. This equilibrium value is $\sim 0.7-0.9$ and corresponds to an
efficiency $\epsilon\sim 0.10-0.20$ \citep[e.g.,][]{Gammie04, Shapiro05,
HBK07}. If the accretion rate of a MBH is less than the Eddington rate by a
factor much larger than $100$ (e.g., $\dot{m}\equiv\dot{\bh}/\dot{\bh}_{\rm
,Edd}\la 10^{-3}$), the MBH may accrete material via the Advection Dominated
Accretion Flow (ADAF) with very low efficiency, $\epsilon \ll 0.1$ (e.g.,
\citealt{NY94}), or via a mode described by the Advection Dominated Inflow and
Outflow scenario (ADIOS, \citealt{BB99}) with most of the accretion material
blown away. The contribution from these very low efficiency modes to the
observational range of the time integral of the QSO/AGN LF is negligible and
MBH growth may also be very inefficient in this low-accretion rate mode. In
this paper, we will not consider this complication but assume that $\epsilon$
is a constant that is neither directly nor indirectly related to the BH mass
$\bh$ and the accretion rate, as $\epsilon$ is probably mainly determined by
the spin of the central BH in the thin-disk accretion mode. (A more detailed
study of the growth of MBHs should simultaneously consider the spin and mass
evolution of MBHs.)

\item If a MBH-disk accretion system accretes material via the Eddington
accretion rate and radiates with luminosity $L_{\rm bol}$, the mass of 
the MBH is
\be
{\bh}_{,\rm Edd}(L_{\rm bol})=
\left(\frac{L_{\rm bol}}{1.26\times 10^{38}\ergs}\right)\msun.
\label{eq:Medd}
\ee

\item The Salpeter timescale is defined as the time for a MBH radiating at the
Eddington luminosity to e-fold in its mass:
\be
\taus\equiv\frac{\bh}{\dot{M}_{\bullet,{\rm Edd}}}=4.5\times 10^7 
\frac{\epsilon}{0.1(1-\epsilon)}\yr.
\ee
If the accretion rate is only a fraction $\lambda$ of the Eddington accretion
rate, then the timescale for a MBH to e-fold its mass is
$\taus'=\lambda^{-1}\taus$.

\end{itemize}

\subsection{Model (a)}\label{subsec:modela}

The mass of MBHs in QSOs may be estimated by using the virial mass
estimator(s), i.e., using the width of broad emission lines and
the empirical relation between the optical luminosities and the sizes of broad
line regions estimated from reverberation mapping studies
(e.g., \citealt{WPM99,Kaspi00,Vestergaard02,Kaspi05}; see also discussion of
uncertainties, e.g., in \citealt{Krolik01}), and hence the Eddington ratio may
be estimated \citep[e.g.,][]{WU02}. Recent studies by \citet{Kollmeier06} on a
sample of QSOs using the virial mass estimator(s) have suggested that the
Eddington ratios ($\dot{m}= L\bol/L\Edd$) in QSOs, may be consistent with a
single value, and the best estimates of the mean value of $\dot{m}$ is around
$10^{-0.6}$ for all redshifts and luminosities. Using a large sample of QSOs
from SDSS, \citet{Shen07} investigate the Eddington ratio distribution in QSOs
over a range of redshifts and luminosities, however, their results show that
the mean value of the Eddington ratio is a function of redshift and luminosity
and it ranges from $10^{-1.1}$ and $10^{-0.6}$.  \citet{Netzer07} also argue
that the $\dot{m}$ distribution is not consistent with a single value and the
conclusion that a single $\dot{m}$ applies to all QSOs/AGNs might be due to
some unknown selection effects.  Ignoring this concern, for the moment, we
assume that all MBHs in QSOs accrete material at a constant normalized rate
$\dot{m}=\lambda$, i.e., $\dot{\bh}=\lambda \dot{\bh}_{\rm ,Edd}$ while the QSO
is ``on''. The luminosity evolution is
\begin{eqnarray}
L_{\rm bol}(M\bhpr,\tau) &=&\lambda L_{\rm Edd}(M\bhpr)\exp\left[
\frac{\tau-\tau\life(M\bhpr)}{\taus'} \right], \nonumber \\
& & \qquad {\rm for\ \ \ } 0<\tau<\tau\life(M\bhpr).
\end{eqnarray}
This model involves three parameters $(\epsilon, \lambda, \xi)$, where
$\xi=\tau\life(M\bhpr)/\taus'=\lambda\tau\life(M\bhpr)/\taus$; and these three
parameters solely determine the growth history of individual MBHs.
For MBHs with present-day mass $M\bhpr$, the probability distribution of 
the nuclear luminosity in their evolutionary history (eq.~\ref{eq:PLMbh}) is
\be
P(L_{\rm bol}|M\bhpr)=\frac{f_Q}{\xi}\frac{1}{L_{\rm bol}},
\label{PLM:ma}
\ee
where
\be
f_Q=\cases{1 & if
$\frac{{\bh}_{,\rm Edd}(L\bol)}{\lambda} \leq M\bhpr\leq \frac{{\bh}_{,\rm Edd}(L\bol)}{\lambda}\exp(\xi)$, \cr 0 & otherwise,}
\ee
and the present-day mass of a MBH is related to its initial mass $M_{\bullet,i}$
at the time of nuclear activity being triggered by
$M\bhpr=M_{\bullet,i}\exp(\xi)$.

For a given set of parameters $(\epsilon, \lambda, \xi)$, we calculate the time
integrals of the XAGN LF, ${\cal T}_{L_{X},\rm XAGN}^{\rm mod}$ (or ${\cal
T}_{M_{\rm B},\rm XAGN}^{\rm mod}$), using equation (\ref{eq:intYband}).  To do
this, the local BHMF is chosen to be the one estimated by using the
$\mbh-\sigma$ relation by \citet{Lauer07a} as the reference BHMF in this paper
(see the solid blue line in the right bottom panel of Fig~\ref{fig:lmf}). For
given BHMF, BCs, and $\lambda$, the normalization of the inferred time
integrals of XAGN LF, i.e., ${\cal T}_{L_{X},\rm XAGN}^{\rm mod}$, is
proportional to $\epsilon/(1-\epsilon)$ through $\tau'_S$, which can vary by a
factor of 10 for the typical range of $\epsilon$, 0.04--0.31. If $\xi$ is
substantially smaller than 1, ${\cal T}_{L_{X},\rm XAGN}^{\rm mod}$ is also
proportional to $\xi$ because the range of the integration limits over the BH
mass is quite small and thus approximately proportional to $\xi$, and the shape
of the inferred time integrals of the QSO LF is determined by the shape of the
local BHMF. In this case, there is some degeneracy between the parameters $\xi$
and $\epsilon$ if $\xi<1$.  However, this degeneracy does not exist if $\xi$ is
substantially larger than 1 (i.e., if the growth of MBHs is dominated by
accretion processes) [which is also true for models (b) and (c) below], as
${\cal T}_{L_{X},\rm XAGN}^{\rm mod}$ is insensitive to $\xi$ at the
high-luminosity end and increases only slowly with increasing $\xi$ at the
low-luminosity end. For example, the predicted ${\cal T}_{L_{X},\rm
XAGN}^{\rm mod}$ for the case of $\xi=2$ (but fixed $\epsilon$ and $\lambda$)
at the low-luminosity end ($L_{X}\la 10^{43}\ergs$) is smaller than that
for the case of $\xi=10$ (with the same $\epsilon$ and $\lambda$) by a factor
of $\sim 1.2-1.3$, and ${\cal T}_{L_{X},\rm XAGN}^{\rm mod}$ for these two
cases are almost the same at the high-luminosity end ($L_{X}\ga 10^{44}
\ergs$).

\begin{figure}
\begin{center}
\includegraphics[width=\hsize,angle=0]{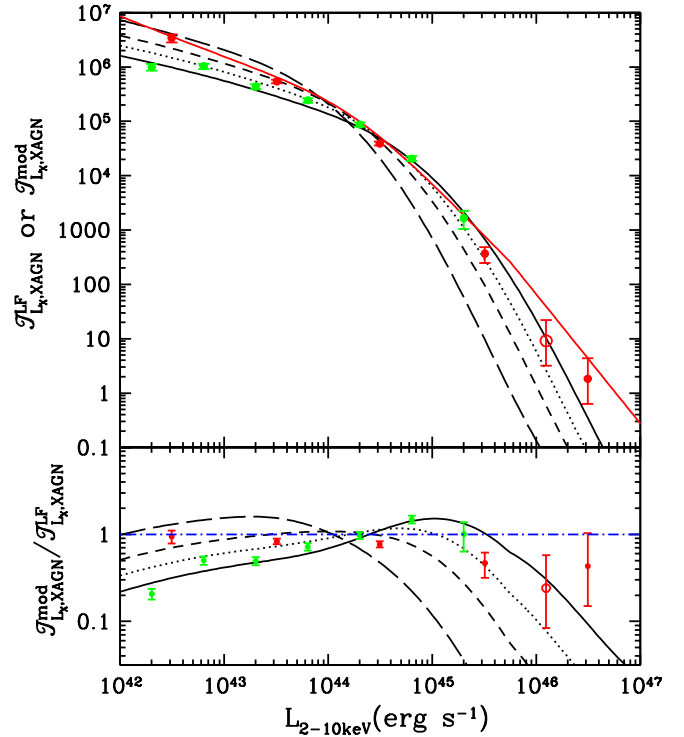}
\caption{
Comparison of the time integral of the X-ray AGN luminosity function (XAGN LF)
and that inferred from the local BHMF by adopting model (a) in
\S~\ref{sec:models}, i.e., assuming that all MBHs accrete material via a fixed
Eddington ratio ($\lambda$). The symbols and the red line are the same as in
Fig.~\ref{fig:XAGNILF}. The black lines represent the inferred time integrals
of the XAGN LF from the local BHMF with $\epsilon=0.14$, $\xi=10$, and
$\lambda=1$ (solid line), 0.5 (dotted line), 0.25 (short-dashed line), and 0.08
(long-dashed line), respectively. The bottom panel shows the inferred time
integrals of the XAGN LF from the BHMF compared to the prediction of the
fitting-formula of XAGN LF obtained by \citet{LaFranca05}. Note that the
estimate of the highest-luminosity point at $L_{X}=10^{46.5}\ergs$ cannot
fit into any model, which might be partly because this point is estimated in
\citet{LaFranca05} from only two AGN with luminosity $\sim 10^{46.1}\ergs$ in
the bin $10^{46}-10^{47}\ergs$. We use these two AGNs to give an estimate on
the space density of AGN at $\sim 10^{46.1}\ergs$ in a bin
$10^{46}-10^{46.2}\ergs$ and show it as  the open circle in this figure.  The
estimates obtained from \citet{LaFranca05} at the low-luminosity end ($L_{\rm
X}\la 10^{43.5}\ergs$, the red points) are adopted as the best one since the
low-luminosity data from \citet{Silverman08} may suffer from selection bias,
while at middle- and    high-luminosities ($10^{43.5}\ergs<L_{X}<10^{46}
\ergs$, the green points) both the estimates from \citet{LaFranca05} and
\citet{Silverman08} are taken into account.  As shown in this Figure, the
inferred time-integrals of XAGN LF cannot match the observations simultaneously
at both the low-luminosity and high-luminosity ends.
}                                                             
\label{fig:msigmma}                                           
\end{center}                                                  
\end{figure}

As shown in Figure~\ref{fig:msigmma}, $\lambda$ should be in the range from 0.5
to 1 in order to match the observations at high luminosity ($L_{\rm
X}>10^{44}\ergs$) with ${\cal T}_{L_{X},\rm XAGN}^{\rm mod}$, while it
should be close to $0.1$ in order to match the observations at lower
luminosities ($L_{X}<10^{43.5}\ergs$).  According to
Figure~\ref{fig:msigmma}, we conclude that the inferred time integrals of the
XAGN LF cannot match the observations at both the low-luminosity and the
high-luminosity ends simultaneously if all MBHs accrete material at a single
$\dot{m}=\lambda$. 

We should note here that ${\cal T}_{L_{X},\rm XAGN}^{\rm mod}$ may well fit the
observations if $\lambda$ is arbitrarily assumed to be an increasing function
of $\mbh$ (cf., the Eddington ratio may be redshift-dependent and thus
mass-dependent since statistically MBHs with larger $\mbh$ formed earlier, see
\citealt{Shankar04,Shankar07}). However, the assumption that all low-mass MBHs
need to accrete material via lower Eddington ratios may be not
realistic/physical because (1) some low-mass MBHs, such as the one in NGC 3079
(with MBH mass $\sim 2\times 10^6\msun$) or NGC 1068 (with MBH mass $\sim
8\times 10^6\msun$), do accrete material with a rate close to the Eddington
limit and have massive accretion disks with mass comparable to the MBH mass
\citep{Kondratko05,LB03}; and (2) there is no clear physical reason for the
low-mass MBHs to accrete material via smaller Eddington ratios compared to
high-mass MBHs if they also obtained their mass mainly from accretion.
Therefore, we do not pursue the possibility that the Eddington ratio is
constant for each AGN with the same MBH mass but an increasing function of the
MBH mass.

\subsection{Model (b)}\label{subsec:modelb}

A more realistic model would be that the growth of MBHs involves two phases
after the nuclear activity is triggered (see the discussions in \citealt{SB92},
\citealt{Blandford03}, and \citealt{YL04a}).  In the first phase, there is
plenty of material to supply the MBH growth; however, MBHs may not be able to
accrete as fast as material fueling allows because the accretion process may be
self-regulated by the Eddington limit. With the decline of the material supply,
the MBH growth enters the second phase and the nuclear luminosity in which the
limiting factor is the fuel supply and accretion rate are expected to decline
to below the Eddington limit.

After the nuclear activity of a MBH is triggered at cosmic time $t_i$, we
assume that the MBH accretes material via the Eddington accretion rate for a
time-period of $\tau\I= \xi\taus$, hence its mass increases to $M\bhtI$ and its
luminosity approaches a peak of $L_{P}(\mbh)=L\Edd(M\bhtI)$ at time
$t\I=t_i +\tau\I$. The nuclear luminosity in this phase increases with time as
\be
{ L}\bol(\tau)=L\Edd(M\bhtI)
\exp\left(\frac{\tau-\tau\I}{\taus}\right),
\qquad 0<\tau<\tau\I,
\label{eq:Lphase1}
\ee
where $\tau=t-t_i$ is the age of the QSO since the nuclear activity was 
triggered.

In the second phase, we assume that the evolution of the nuclear luminosity
(or accretion rate) declines exponentially as (e.g.,\ \citealt{HNR98,HL98}):
\begin{eqnarray}
& & { L}\bol(\mbh,\tau)= \nonumber \\
& & \cases{L\Edd(M\bhtI)\exp\left(-\frac{\tau-\tau\I}{\tau\D}
\right), &  for $\tau\I\le\tau\le\tau\I+\eta\tau\D$, \cr
0, & for $\tau>\tau\I+\eta\tau\D$,} \nonumber \\
& & 
\label{eq:Lphase2}
\end{eqnarray}
where $\tau\D=\zeta\taus$ is the characteristic decay timescale of the
nuclear luminosity. We assume that QSOs become quiescent when the nuclear
luminosity declines by a factor of $\exp(-\eta)$ compared to the peak
luminosity $L\Edd(M\bhtI)$, so there is a cutoff of the nuclear luminosity at
$\tau\life=\tau\I+\eta\tau\D=(\xi+\eta\zeta)\taus$ in equation
(\ref{eq:Lphase2}).  The factor $\eta$ is set to $-\ln(10^{-3})=6.9$ here,
since after decreasing by a factor of $10^{-3}$ in accretion rate, the
accretion mode may change from the efficient thin-disk accretion to the
inefficient advection dominated accretion modes and the nuclear luminosity of
MBHs even with a high mass $\sim 10^9\msun$ will become fainter than the
luminosity range ($M_{\rm B}\la -20$ or $L_{X}\la 10^{42}\ergs$) of interest in this
paper.  With the assumption that all QSOs are quenched at present (i.e.,
$t_0-t_i-\tau\I \gg \tau\D$), the MBH mass at the present day is
\be
M\bhpr\simeq \left(1+\zeta\right)M\bhtI=(1+\zeta)\exp(\xi)M_{\bullet,i}.
\label{eq:Mbhprb}
\ee
This model reduces to model (a) with $\lambda=1$ if the second phase is not
significant.

For MBHs with present-day mass $M\bhpr$, the probability distribution of the
nuclear bolometric luminosity in their evolutionary history
(eq.~\ref{eq:PLMbh}) is
\be
P(L\bol|M\bhpr)
=\frac{f\I+\zeta f\D}{\xi+\eta\zeta}\frac{1}{L\bol},
\qquad
\label{eq:PLMQSOb}
\ee
where
\be
f\I= \nonumber \\
\cases{1 & if $(1+\zeta){\bh}_{,\rm Edd}(L_{\rm bol})\leq M\bhpr\leq$ \cr   
& \qquad $(1+\zeta)\exp(\xi){\bh}_{,\rm Edd}(L_{\rm bol})$, \cr
 0 & otherwise,
}
\ee
and
\be
f\D=\cases{1 & if
$(1+\zeta){\bh}_{,\rm Edd}(L_{\rm bol}) \leq M\bhpr \leq$\cr
& \qquad  $10^3 (1+\zeta){\bh}_{,\rm Edd}(L_{\rm bol})$, 
\cr
0 & otherwise.
}
\ee
In model (b), we also have three parameters $(\epsilon,\xi,\zeta)$ to
be constrained below.

\begin{figure}                                                
\begin{center}                                                
\includegraphics[width=\hsize,angle=0]{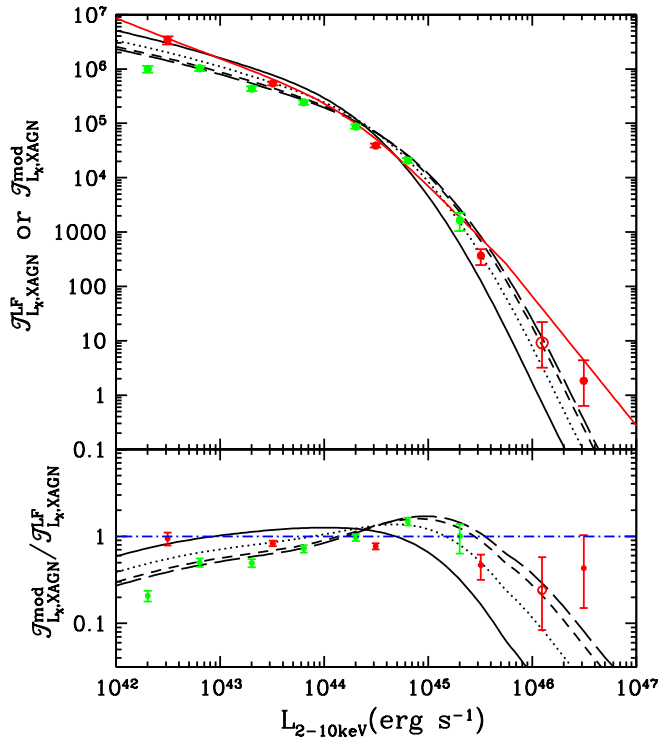}
\caption{                                                     
Similar to Fig.~\ref{fig:msigmma}, but adopting model (b) in  
\S~\ref{sec:models}, i.e., an initial accretion phase with rate set by the  
Eddington limit, followed by a phase with an exponentially declining accretion
rate.  The black lines represent the inferred time integrals of XAGN LF from
the local BHMF with $\epsilon=0.16$, $\xi=10$, and $\zeta=0.1$ (long-dashed 
line), 0.3 (short-dashed line), 1 (dotted line) and 3 (solid line),
respectively.  As shown in this Figure, the inferred time-integrals of XAGN LF
cannot match the observations simultaneously at both the low-luminosity and 
high-luminosity ends.                                         
}                                                             
\label{fig:msigmmb}                                           
\end{center}                                                  
\end{figure}

For any given set of parameters ($\epsilon$, $\xi$, $\zeta$), we calculate the
time integrals of the XAGN LF, ${\cal T}^{\rm mod}_{L_{X},\rm XAGN}$. Our
calculations show that the dependence of ${\cal T}^{\rm mod}_{L_{X},\rm
XAGN}$ on parameters $\epsilon$ or $\xi$ for a given $\zeta$ is similar to that
in model (a). For given $\epsilon$ and $\xi$, a larger $\zeta$ corresponds to
smaller ${\cal T}^{\rm mod}_{L_{X},\rm XAGN}$ at higher luminosities but
larger ${\cal T}^{\rm mod}_{L_{X},\rm XAGN}$ at lower luminosities. As
shown in Figure~\ref{fig:msigmmb}, $\zeta$ must be around or smaller than
$0.1-0.3$ to match ${\cal T}^{\rm mod}_{L_{X},\rm XAGN}$ with observations
at the high-luminosity end ($L_{X}\ga 10^{44.5}\ergs$), but $\zeta$ should
be larger than $1$ to match ${\cal T}^{\rm mod}_{L_{X},\rm XAGN}$ with
observations at the low-luminosity end ($L_{X} \la 10^{43}\ergs$). It is
unlikely that ${\cal T}^{\rm mod}_{L_{X},\rm XAGN}$ inferred from any
single $\zeta$ with any fixed ($\epsilon$, $\xi$) can match observations
simultaneously at both the high- and low-luminosity end.

As discussed in model (a), $\xi$ and $\zeta$ are not necessarily constants in
model (b) but may be functions of $M\bhpr$; or alternatively the ratio of the
MBH mass $M\bhtI$ at the peak luminosity to the final MBH mass $\mbh$ may be a
slowly increasing function of $\mbh$ as proposed by \citet{Hopkins06}. The
dependence of $\xi$ and $\zeta$ on $M\bhpr$ would be related to the assembly
history of each MBH and the distribution of seed BHs, which are poorly known.
In model (b), it is possible that ${\cal T}^{\rm mod}_{L_{X},\rm XAGN}$ can match
observations at both the high- and low-luminosity ends if $\zeta$ decreases
with increasing $\mbh$. But we will not go further to make this fit, for the
same reasons given at the end of \S~\ref{subsec:modela}. 

\subsection{Model (c)}\label{subsec:modelc}

The accretion rates in the second phase of QSOs, in which the luminosity
decays, may be ultimately determined by the evolution of the viscous accretion
disk itself rather than galactic-scale dynamical disturbances. The disk
accretion evolution may follow a self-similar solution
\citep[e.g.,][]{Pringle81, LP87, CLG90, Pringle91}, i.e., the accretion rate
declines as a power-law of the QSO age ($\dot{\bh}\propto \tau^{-\gamma}$),
where the slope $\gamma$ may be determined by the opacity law. The value of
$\gamma$ also depends on the binarity of the MBH surrounded by the accretion
disk, for instance, $\gamma\sim 1.2-1.3$ for the evolution of a disk around a
single MBH \citep[e.g.,][]{CLG90}, while $\gamma\sim 2.5-3.3$ for a disk
truncated by an outer secondary MBH \citep[e.g.,][]{LS00}.  For the binary MBH
system, however, the secondary MBH embedded in a disk surrounding the primary
MBH may migrate inward and may merge with the primary MBH on a time-scale of
$10^7$~yr \citep[e.g.,][]{AN02, Escala04, Escala05}, and thus the evolution of
disk accretion associated with the binary MBH system ($\gamma\sim 2.5-3.3$) may
not be sustained for a period substantially longer than $10^7$~yr. The observed
accretion rate distribution in local AGNs is found to be consistent with the
self-similar evolution around a single MBH and $\gamma\simeq 1.26\pm 0.1$
\citep[][see also \citealt{KP07}]{YLK05}. In this paper we neglect the
complications in the evolution of disk accretion due to possible binary MBHs,
and assume $\gamma\sim 1.2-1.3$. Below we introduce model (c) which is similar
to model (b) but the nuclear luminosity in the second phase declines with time
as a power law:
\begin{eqnarray}
& & { L}\bol(\mbh,\tau)= \nonumber \\
& & \cases{L\Edd(M\bhtI)\left(\frac{\tau+\tau\D-\tau\I}{\tau\D}
\right)^{-\gamma}, 
& for $\tau\I\le\tau\le\tau\I+\eta\tau\D$, \cr
0, & for $\tau>\tau\I+\eta\tau\D$,} \nonumber \\
\label{eq:Lphase2c}
\end{eqnarray}
where $\tau\D=\zeta\taus$ is the transition timescale from the first to
the second phase. As in
model (b), we assume that QSOs become quiescent when the nuclear
luminosity declines by a factor of $10^{3}$ compared to the peak luminosity
$L\Edd(M\bhtI)$ and afterwards the growth of MBHs is not significant. Thus
$\eta=10^{3/\gamma}-1$ in this model. The MBH mass at a time $\tau$ after 
the nuclear activity was triggered is
\be
\bh^{\tau}=M\bhtI \exp\left(\frac{\tau-\tau\I}{\taus}\right),
\label{eq:mbhtau1}
\ee
in the first phase, and is 
\be
\bh^{\tau}=M\bhtI\left[1+\frac{\zeta}{\gamma-1}\left(1-
(\frac{\tau+\tau\D-\tau\I}{\tau\D})^{1-\gamma}\right)\right]
\label{eq:mbhtau2}
\ee
in the declining phase.  The present-day mass of a MBH is
\be
M\bhpr\simeq \chi M\bhtI= \chi \exp(\xi)M_{\bullet,i},
\label{eq:Mbhprc}
\ee 
where $\chi=1+\frac{1-10^{3(1-\gamma)/\gamma}}{\gamma-1}\zeta$.
The slope can be $\gamma\sim 1.2$, $1.3$ and thus
$M\bhpr\sim (1+3.41\zeta)M\bhtI$, and $(1+2.66\zeta)M\bhtI$, respectively.

In model (c), for MBHs with present-day mass $M\bhpr$ the probability
distribution of the nuclear bolometric luminosity in their evolutionary history
(eq.~\ref{eq:PLMbh}) is
\be
P(L_{\rm bol}|M\bhpr)
=\frac{f\I+f\D\frac{\zeta}{\gamma}
\left(\frac{L\Edd(M\bhtI)}{L\bol}\right)^{1/\gamma}
}{\xi+\eta\zeta}\frac{1}{L\bol},
\qquad
\label{eq:PLMQSOc}
\ee
where
\be
f\I=\cases{1 & if $\chi {\bh}_{,\rm Edd}(L_{\rm bol}) \le M\bhpr$ \cr
             & \qquad $\leq \chi\exp(\xi) {\bh}_{,\rm Edd}(L_{\rm bol})$
\cr
0 & otherwise,
}
\ee
and
\be
f\D=\cases{1 & if
$\chi {\bh}_{,\rm Edd}(L\bol) \le M\bhpr \leq  10^3 \chi {\bh}_{,\rm Edd}(L\bol)$,
\cr
0 & otherwise.
}
\ee
Besides the three parameters $(\epsilon,\xi,\zeta)$ involved in model (c), an
additional parameter $\gamma$ is also involved, but $\gamma$ is fixed by
assumption to be
1.2--1.3 here, if not otherwise specified, according to theoretical models on
the long-term evolution of viscous disk
\citep[e.g.,][]{Pringle81,LP87,CLG90,Pringle91} and recent observational
constraints \citep[e.g.,][see also \citealt{KP07}]{YLK05}.

\begin{figure}
\begin{center}
\includegraphics[width=\hsize,angle=0]{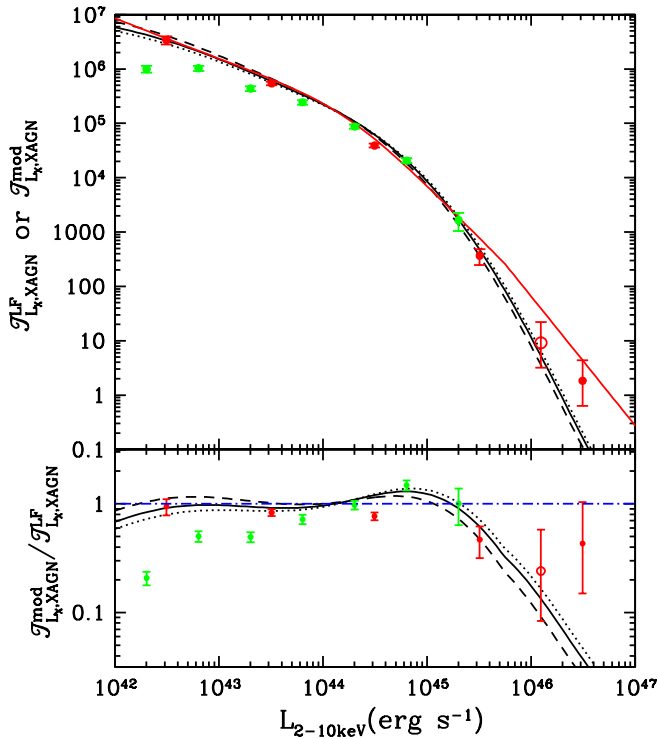}
\caption{
Similar to Fig.~\ref{fig:msigmma}, but adopting model (c) in
\S~\ref{sec:models}, i.e., an initial accretion phase with rates set by the
Eddington limit, followed by a phase with power-law declining accretion rates,
as might be set by self-similar long-term evolution of disk accretion. The
black lines represent the inferred time integrals of XAGN LF from the local
BHMF with $\epsilon=0.16$, $\xi=10$, and $\zeta=0.15$ (dotted line), 0.20
(solid line) and 0.30 (dashed line), respectively. This Figure shows that the
time-integral of XAGN LF inferred from the local BHMF can well  match
observations simultaneously at both the low-luminosity and high-luminosity ends
with suitable parameters. We take the model (c) with parameters ($\epsilon$,
$\xi$, $\zeta$)=(0.16, 10, 0.20) as the reference model in this paper.
}
\label{fig:msigmmc}
\end{center}
\end{figure}

For any given $\zeta$ in this model, the dependence of ${\cal T}^{\rm
mod}_{L_{X},\rm XAGN}$ on the parameters $\epsilon$ and $\xi$ is similar to
that in models (a) and (b). For given $\epsilon$ and $\xi$, a larger $\zeta$ is
responsible for a smaller ${\cal T}^{\rm mod}_{L_{X},\rm XAGN}$ at higher
luminosities but a larger ${\cal T}^{\rm mod}_{L_{X},\rm XAGN}$ at lower
luminosities because a larger $\zeta$ means that a larger fraction of the mass
of MBHs is accreted via the second phase with Eddington ratios (substantially)
smaller than 1. Although the growth history of both low-mass MBHs and high-mass
MBHs is assumed the same in this model for fixed parameters ($\epsilon$, $\xi$,
$\zeta$), apparently there are more objects with low Eddington ratios at the
low-luminosity end but few objects with low Eddington ratios at the
high-luminosity end. Detailed investigation of the Eddington ratio distribution
inferred from this model is discussed in \S~\ref{sec:accrete}.  As shown in
Figure~\ref{fig:msigmmc}, ${\cal T}^{\rm mod}_{L_{X},\rm XAGN}$ can match
observations very well if $\zeta \sim 0.15-0.3$ provided that $\epsilon=0.16$,
$\xi\ga 2$ and $\gamma \sim 1.2-1.3$. With the parameter $\zeta\sim 0.15-0.3$
and $\gamma\sim 1.2-1.3$, the mass growth of MBHs at the accretion stage with
Eddington ratio $\dot{m}<1$ ( or $\dot{m}\la 0.1$) is roughly a fraction $\sim
0.2-0.5$ (or $\sim 0.1-0.3$) of its final mass $M\bhpr$, and this is compatible
with the assumption that the disk mass is substantially less than the central
MBH in the long-term evolution of disk accretion model
\citep[e.g.,][]{Pringle81, LP87, CLG90, Pringle91}; and MBHs obtained majority
of their mass ($\ga 80\%$) via a rate close to the Eddington limit ($\dot{m}\ga
0.1$). With these parameters, we have $\tau\life(\mbh)\sim
\taus[\xi+(10^{3/\gamma}-1) \zeta] \sim (3-6)\times 10^9$~yr, and the period
for MBH-accretion disk systems radiating at luminosities larger than $10\%$ of
its peak luminosity, thus roughly $\dot{m}\ga 0.1$ (or $\dot{m}\ga0.01$), is
only about $(3-4)\taus\sim (2-3)\times 10^8$~yr (or $\sim  10^9$~yr).  Model
(c) is based on detailed considerations of the evolution of disk accretion and
appears to fit observations much better than models (a) and (b). Therefore,
this model with three parameters ($\epsilon$, $\xi$, $\zeta$)=($0.16$, $10$,
$0.20$) is set as the reference model in this paper.

Considering of the uncertainty in the $\mbh-\sigma$ relation, the velocity
dispersion distribution function and the time integral of XAGN LF, the error in
the best-matched parameter $\epsilon$ is $\delta \epsilon\sim 0.04$. Note also
that Compton-thick objects may be still missed in the hard X-ray surveys by
\citet{LaFranca05}. The fraction of Compton-thick objects should not be larger
than $30$\% according to the X-ray background synthesis model
\citep[e.g.,][]{MH07}, and this would add additional uncertainty at most
$^{+0.05}_{-0}$ to $\epsilon$. To match the time integral of XAGN LF with
the local BHMF, the efficiency $\epsilon$ is required to $\simeq 0.16\pm 0.04
^{+0.05}_{-0}$, and this range of $\epsilon$ is fully consistent with
theoretical expectations $\epsilon\sim 0.10-0.20$ \citep[e.g.,][]{Gammie04,
Shapiro05, HBK07}. The range of $\epsilon$ ($\sim 0.12-0.25$) constrained above
corresponds to the spin parameter $a$ in the range from $\sim 0.8$ to $0.99$
as the value of $\epsilon$ is mainly determined by $a$ with only an order of
$10-20\%$ or less
uncertainty (e.g., \citealt{Nobleetal}), which suggests that most MBHs in QSOs are indeed rapidly
rotating Kerr BHs.  It is worth to note that if we choose $\epsilon=0.21$, the
time integral of XAGN LF inferred from the local BHMF with parameters ($\xi
\sim 1$, $\zeta\sim 0.2$) can still match the observations well, but the time
integral of XAGN LF at luminosities $L_{X}\la 10^{45}\ergs$ is overpredicted by
$\sim 40\%$ if ($\xi \sim 10$, $\zeta\sim 0.2$) and the overpredicted part can be
accounted for by the additional contribution from Compton-thick AGNs. Previous
estimates
of the efficiency include $\epsilon\ga 0.15$ \citep{Elvis02}, $\epsilon\ga 0.1$
\citep{YT02,YL04a}, $\epsilon\sim 0.04-0.16$ \citep{Marconi04}, $\epsilon\sim
0.30-0.35$ \citep{Wang06}, and $\epsilon\sim 0.06-0.11$ \citep{Shankar04,
Shankar07}. 

The relatively high efficiency constrained above suggests that the majority of
QSOs should not accrete material via the chaotic accretion scenario proposed by
\citet{KP07a} to explain the rapid growth of MBHs in those QSOs at $z>6$, in
which MBHs spin down because of counter-alignments of their spin axes with
accretion disk angular momenta and thus the efficiency reduces to a low value,
close to the efficiency for Schwarzschild BHs.

Note that there are some uncertainties in the intrinsic scatter in the
$\mbh-\sigma$ relation, which may mainly introduce some uncertainties to the
parameter $\zeta$.  A larger intrinsic scatter corresponds to more MBHs at the
high-mass end and thus allows a larger $\zeta$. But the uncertainties in
$\zeta$ introduced by the uncertainties in the intrinsic scatter is not
significant if this uncertainty in the scatter is less than $0.1$~dex (for
example, it is about $0.06$~dex in \citealt{Tundo07}). 

In the above models, we adopt the local BHMF estimated from the $\mbh-\sigma$
relation and the velocity-dispersion distribution function. Arguably the
$\mbh-L\bulge$ relation may be favored, at least for the most massive galaxies
(see \citealt{Lauer07a}, but \citealt{Batcheldor07} and \citealt{Graham08}).
\footnote{Although the $\mbh-L\bulge$ relation may be favored according to the
observations for BCGs, which primarily infer more massive BHs in BCGs compared
with that inferred from the $\mbh-\sigma$ relation, the most massive BHs may
mostly be found in galaxies less massive than BCGs if the intrinsic scatter in
the $\mbh-\sigma$ (or $\mbh-L\bulge$) relation is significant (e.g., $\ga
0.3$~dex). } If we adopt the local BHMF estimated from the $\mbh-L\bulge$
relation and the galaxy luminosity function without correction of the bias as
discussed in \S~\ref{sec:BHMF}, the time integral of XAGN LF can be matched by
${\cal T}^{\rm mod}_{L_{X},\rm XAGN}$ inferred from the local BHMF if
$\xi=10$, $\zeta\sim 0.15-0.3$ but $\epsilon \simeq 0.08\pm
0.02^{+0.03}_{-0}$ (correspondingly the spin parameter $a$ is in the range from $0.1$ to $0.8$); and therefore the QSO lifetime constrained here is
smaller than that constrained by the local BHMF obtained from the $\mbh-\sigma$
relation by a factor $\sim 2$.  This is primarily due to the fact that the
shape of the local BHMF estimated from the $\mbh-L\bulge$ relation is similar
to that estimated from the $\mbh-\sigma$ relation except that the normalization
differs by a factor close to $2$ (see discussions in \S~\ref{sec:BHMF} and the
bottom left panel of Fig.~\ref{fig:lmf}).

\section{Clues on the luminosity dependence of the obscuration
fraction of AGNs}\label{sec:obscuration}

\begin{figure}
\begin{center}
\includegraphics[width=\hsize,angle=0]{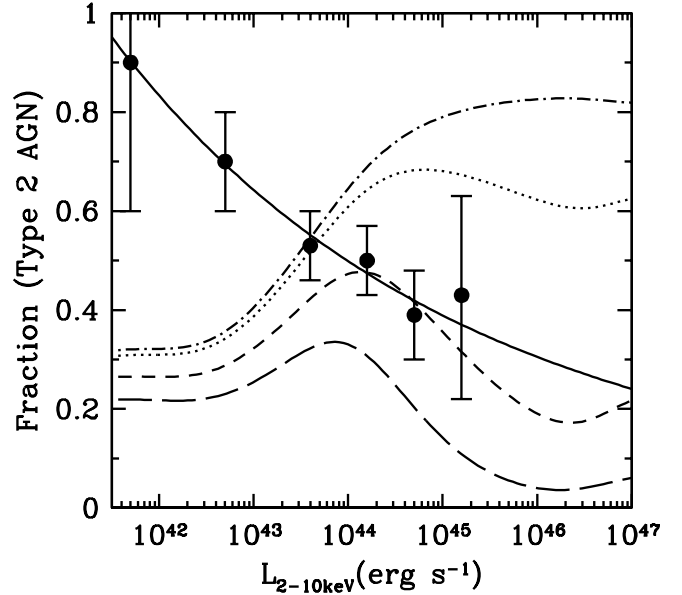}
\caption{
Fraction of type 2 AGNs as a function of X-ray luminosity. Circles with error
bars are the observed fraction of obscured AGNs and the solid line is the
best-fit model to the red circles given by \citet{Akylas06}.  Other lines
represent the fraction of obscured AGNs expected from the reference model if we
assume that all AGNs in the first rapid growth phase with $\tau$ in the ranges
$[0, \tau_P-\taus]$ (long-dashed line), $[0,\tau_P-0.5\taus]$ (short-dashed
line), $[0,\tau_P-0.1\taus]$ (dotted line), or $[0,\tau_P]$ (dot-dashed line),
are obscured, while other AGNs including those in the second (declining) phase
are all unobscured. The dot-dashed line seems to be consistent with
observations at high luminosities ($L_{X}\ga 10^{43.5}\ergs$), but all blue
lines are significantly lower than the observations at the low luminosity end,
which suggests that the obscuration of AGNs cannot be solely an evolutionary
effect arising from their individual evolution after nuclear activity is
triggered.  
}
\label{fig:obsc}
\end{center}
\end{figure}

Many X-ray studies have shown that the fraction of type 2 (or heavily obscured)
AGNs decreases with increasing X-ray luminosity \citep[][as shown by the red
solid line and data points in Fig.~\ref{fig:obsc}]{Ueda03, Akylas06, MH07},
though there are still some uncertainties about whether this relation is real
or just a selection effect \citep[e.g.,][]{LaFranca05, Treister06, Akylas06,
Tozzi06}. This relation may be explained in the current evolutionary model for
QSOs/AGNs \citep[e.g.,][]{Hopkins05}, i.e., QSOs/AGNs in their rapid growth
phase are moderately luminous and more likely to be heavily obscured, and as
the AGN luminosity increases the UV-X-ray photons emitted from the QSOs/AGNs
destroy the surrounding absorbing material and the QSOs/AGNs become unobscured. 

In \S~\ref{sec:models}, we have shown that the time integrals of the X-ray LF
estimated from observations can be well matched by those inferred from the
local BHMF within the reference model of the growth of individual MBHs.  In the
reference model [i.e., model (c) with parameters $(\epsilon,\xi,\zeta)=
(0.16,10,0.20)$], however, type 1 and type 2 AGNs are not distinguished.  In
this section, we use a simple toy model to check whether the dependence of the
fraction of type 2 AGNs on the X-ray luminosity can be really due to
evolutionary effect described in the preceding paragraph. 

In our toy model, we assume that those QSOs/AGNs at their early rapid growth
stage are all obscured while those in their late evolutionary stage are all
un-obscured.  The inferred fraction of obscured QSOs/AGNs is shown in
Figure~\ref{fig:obsc} if all QSOs/AGNs in the first rapid growth phase (i.e.,
the Eddington accretion stage) with $\tau$ in the ranges $[0, \tau_P-\taus]$,
$[0,\tau_P-0.5\taus]$, $[0,\tau_P-0.1\taus]$, or $[0,\tau_P]$ are assumed to be
obscured (shown as long-dashed, short-dashed, dotted and dot-dashed lines in
Fig.~\ref{fig:obsc}).  The observations of the fraction of type 2 AGNs as a
function of the X-ray luminosity is also shown in Figure~\ref{fig:obsc}.
Although the short-dashed line in Figure~\ref{fig:obsc} is not inconsistent
with the observational trend at the high luminosity end ($L_{X}\ga
10^{43.5}\ergs$), clearly the trend of the dependence of the fraction of type 2
AGNs on the X-ray luminosity implied by these toy models is in contradiction
with observations at the low-luminosity end, which suggests that the
obscuration of AGNs cannot be solely an evolutionary effect arising from their
individual evolution after their nuclear activities are triggered. Some other
effect, such as those introduced by the receding torus model
\citep[e.g.,][]{Lawrence91,Simpson05} in which the opening angle of the torus
is smaller in less luminous QSOs/AGNs, should be responsible for the larger
fraction of type 2 AGNs at low luminosity.

\begin{figure}
\begin{center}
\includegraphics[width=0.95\hsize,angle=0]{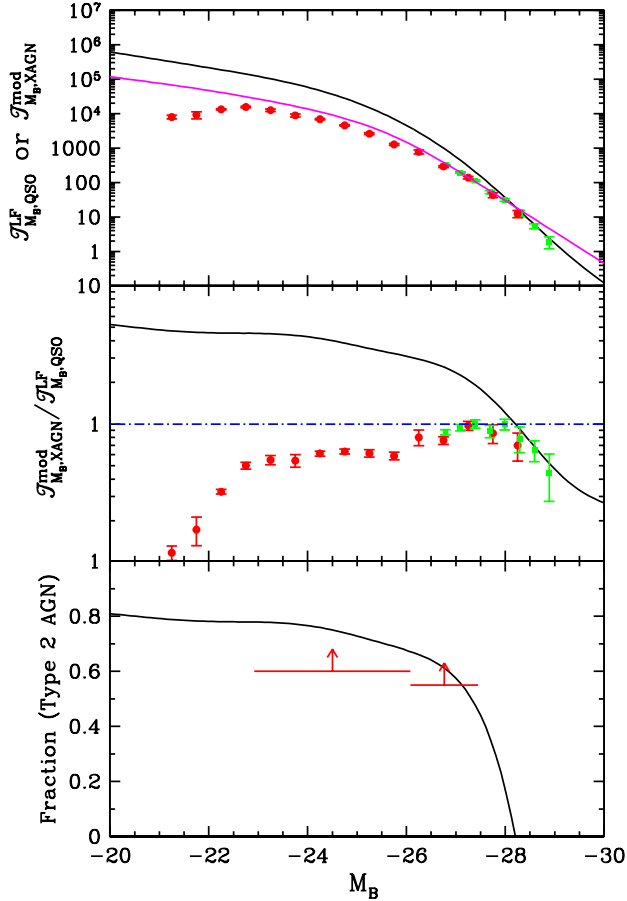}
\caption{
Model (c) for the $B$-band. The upper panel represents the comparison of the
time integral of the $B$-band QSO LF and that inferred from the local BHMF by
adopting the reference model. The black line represents the value inferred from
the local BHMF by using the reference model, i.e., model (c) with parameters
$(\epsilon,\xi,\zeta)=(0.16,10,0.20)$ (similarly in middle and bottom panels).
The magenta line represents the values obtained from the fitting-formula of the
QSO LF obtained by \citet{Richards05}. The points have similar meanings as in
Fig.~\ref{fig:QSOILFadd}. The middle panel shows the ratio of the values
represented by the black line and the points in the upper panel to the values
represented by the magenta line, i.e., the values in units of that inferred
from the fitting-formula of the QSO LF. The bottom panel shows the fraction of
optically obscured AGNs inferred from the reference model. The upper arrows
show the lower limit of the fraction of optically obscured AGN at $z<0.3$
($M_{\rm B}\sim-22.9$ ---$-26.1$) and $0.3<z<0.83$ ($M_{\rm B}\sim-26.1$
---$-27.5$), respectively \citep{Reyes08}, which are consistent with the
prediction of the reference model in this paper. The range of $M_{\rm B}$ is
converted from the luminosity range of [OIII] 5008\AA~line [using
$\log(L[OIII]/L_{\sun})=-0.38M_{2400}-0.62$ in Fig.~11 in \citet{Reyes08},
where $M_{2400}$ is the absolute magnitude at 2400\AA~in the rest frame, and
$M_{\rm B}\sim M_{2400}+0.13$ by assuming a canonical optical spectral slope
($\sim 0.5$) of QSOs]. 
}
\label{fig:mcolf}
\end{center}
\end{figure}

Comparison of the time-integral of the QSO LF in the optical band inferred from
the local BHMF with that from observation will also provide information on the
fraction of obscured QSOs in the optical band.  In the upper panel of
Figure~\ref{fig:mcolf}, the black line represents the inferred time integrals
of the QSO LF in the $B$ band for the reference model.  Correspondingly, the
inferred time integral in units of the time integral obtained from the QSO LF
given by \citet{Richards05} is shown in the middle panel.  The reference model
well matches the observations at $M_{\rm B}\la -28$ but predicts more QSOs than
those observed at magnitude $M_{\rm B}\ga -28$, which suggests that there exist
a larger fraction of optically obscured QSOs/AGNs and this fraction is shown in
the bottom panel. The dependence of the optically obscured QSO fraction on the
luminosity in the range $-27\la M_{\rm B} \la -20$ is much weaker compared to
that in the X-ray band as shown in Figure~\ref{fig:obsc}.  As we can see from
the bottom panel, the fraction of optically obscured QSOs/AGNs can be as high
as $80\%$ at $M_{\rm B}\sim-20$ ---$-23$ and slightly decreases to $60\%$ at
$M_{\rm B}\sim -27$. The fraction of $\sim 80\%$ at $M_{\rm B}\sim-20$ ---$-23$
is consistent with the observations that the ratio of Seyfert 2 galaxies to
Seyfert 1 galaxies is about 4:1 in the nearby universe.  The fraction of $\sim
60-70\%$ at $M_{\rm B}\sim-24$ ---$-27$ is consistent with the latest estimates
from \citet{Reyes08} as indicated by the two lower limits, which are converted
from the fractions at the [OIII] 5008\AA ~luminosity measured in
\citet{Reyes08} to the fractions at the $B$-band magnitude, and this consistence
supports the constraints on the growth of MBHs obtained above by applying the
extended So{\l}tan argument to the X-ray data. At higher luminosities, $M_{\rm
B}<-28$, the fraction of optically obscured QSOs sharply decreases to 0, which
may be not genuine but due to effects of uncertainties in the BC at the
high-luminosity end or the local BHMF at the high-mass end. 

\section{The Eddington-ratio distribution in QSOs}
\label{sec:accrete}

Observational determination of the Eddington ratio distribution in QSOs can put
additional constraints on MBH growth. These are independent of, but should be
consistent with, the constraints obtained above from the extended So{\l}tan
argument. Recent observational advances allow us  to seriously estimate the
Eddington ratio distribution in large samples of QSOs. For example, using the
virial mass estimators \citet{Kollmeier06} and \citet{Shen07} have shown that
the logarithm of Eddington-ratio distribution in high-luminosity QSOs resembles
a Gaussian distribution with mean around $10^{-0.6}$ to $10^{-1.1}$ and width
typically of 0.3~dex (see also \citealt{Netzer07}), which may suggest that MBHs
obtain most of their mass through accretion with a rate close to the Eddington
limit. In this section, we check whether the luminosity evolution of individual
QSOs constrained above is consistent with the observational Eddington ratio
distribution.

\begin{figure} 
\begin{center}
\includegraphics[width=\hsize,angle=0]{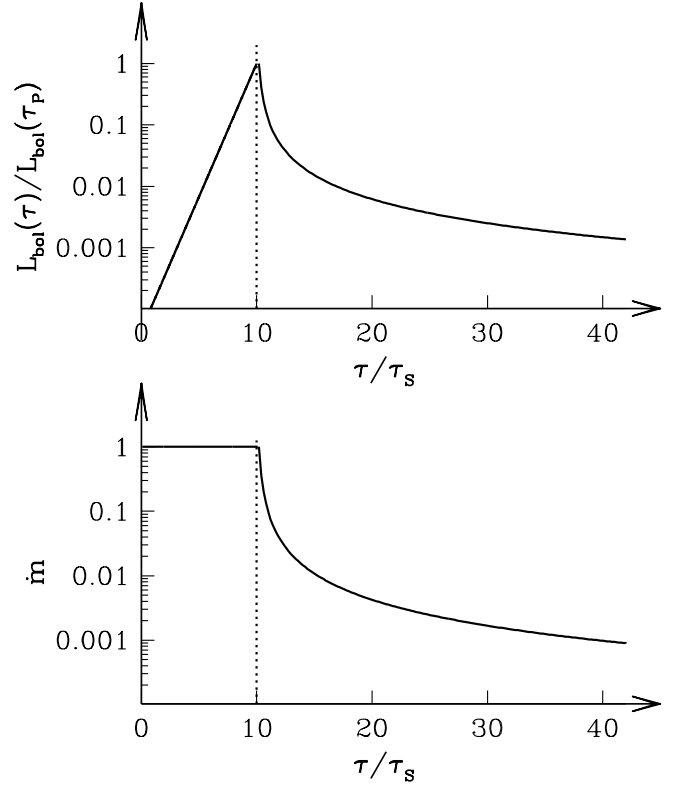}
\caption{The luminosity and Eddington ratio (i.e., the accretion rate in
units of the Eddington rate) evolution curves in the reference model, i.e.,
model (c) with parameters ($\epsilon$, $\xi$, $\zeta$)=($0.16$, $10$,
$0.20$). In the initial rapid accretion phase, the luminosity of a QSO
exponentially increases and the Eddington ratio is a constant
($\sim 1$) due to the self-regulation of the disk accretion when the
accretion material is over-supplied, and both the luminosity and Eddington
ratio of the QSO are followed by a rapid power-law-like decline due to
the exhaustion of fuel and the self-similar evolution of disk accretion
(see also \citealt{YLK05}).
}
\label{fig:lc}
\end{center}
\end{figure}

In \S~\ref{sec:models}, we have shown that the time-integrals of XAGN LF
inferred from the local BHMF can be well matched to the observations using
model (c) with parameters ($\epsilon$, $\xi$, $\zeta$)=(0.16, 10, 0.20) for the
luminosity evolution of individual QSOs. In this reference model, the
luminosity evolution and correspondingly the Eddington ratio evolution of a QSO
are illustrated in Figure~\ref{fig:lc}. As shown in the upper panel of
Figure~\ref{fig:lc}, the luminosity of a QSO exponentially increases to its
peak luminosity with the Eddington rate set by the self-regulation of disk
accretion when the fuel is over-supplied, and then decays with time as a
power-law set by the self-similar evolution of disk accretion when the fuel is
substantially under-supplied. The period for the QSO to have luminosity larger
than $10\%$ of its peak luminosity is only a few times the Salpeter timescale.
Correspondingly the Eddington ratio of the QSO is initially about $1$ and then
also decays with time approximately as a power-law (the bottom panel of
Fig.~\ref{fig:lc}). The timescale for $\dot{m}$ declining from $1$ to $0.1$ is
relatively short compared to the Salpeter timescale, and those QSOs around its
peak luminosity should mainly accrete material via Eddington ratio close to
$1$. As mentioned in \S~\ref{sec:intro}, the QSO LF at different redshifts
involves the dependence on both the nuclear activity triggering rate
${\cal G}(z;\mbh)$ and the luminosity evolution of individual QSOs ${\cal
L}(\tau;\mbh)$ after their nuclear activity being triggered, so does the QSO
Eddington-ratio distribution at different redshifts.  Below we define a
`time-integrated' Eddington-ratio distribution in QSOs, which only involves the
accretion-rate evolution of individual QSOs.

With a luminosity evolution model, the true Eddington ratio ($\dot{m}^{\rm r}$)
distribution at a fixed bolometric luminosity $L\bol$ for a MBH with
present-day or final mass $\mbh$ is
\be
P(\dot{m}^{\rm r}|L\bol,\mbh)d\dot{m}^{\rm r}
=\delta(\dot{m}^{\rm r}-\dot{m}_0)d\dot{m}^{\rm r},
\label{eq:pmdotindi}
\ee
where $\dot{m}_0=1$ if the luminosity evolution is in the first rapidly
increasing phase, $\dot{m}_0=M_{\bullet,\rm Edd}(L\bol)/\bh^{\tau}$ if in the
decline phase, and $\bh^{\tau}$ is the mass of the MBH in a QSO with bolometric
luminosity $L\bol$ and with its final mass $\mbh$. The $\bh^{\tau}$ can be
directly obtained from equation (\ref{eq:mbhtau1})  or (\ref{eq:mbhtau2}) for
fixed $L\bol$ and $\mbh$. With given $P(\dot{m}^{\rm r}|L\bol,\mbh)$, the
(`time-integrated') probability distribution of the Eddington ratios among QSOs
at a given $L\bol$ can be defined by
\begin{eqnarray}
& & P(\dot{m}^{\rm r}|L\bol)= \int n(\mbh)\tau\life(\mbh)\times \nonumber \\
& & P(L\bol|\mbh)P(\dot{m}^{\rm r}|L\bol,\mbh)d\mbh\times \nonumber \\
& & \left[\int n(\mbh) \tau\life(\mbh)P(L\bol|\mbh) d\mbh\right]^{-1}
\label{eq:probmdotr}
\end{eqnarray}
Note that the denominator in the above equation is just the time integral of
the QSO LF.

\begin{figure}                                                
\begin{center}                                                
\includegraphics[width=\hsize,angle=0]{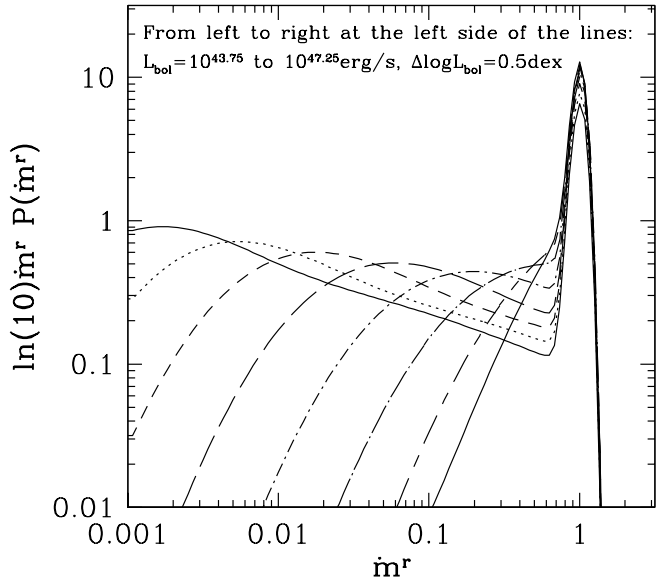}
\caption{
The underlying `time-integrated' Eddington ratio distribution among QSOs at a
given bolometric luminosity inferred from the reference model     in
\S~\ref{sec:models}. From left to right at the low-Eddington ratio end,  the
lines represent $P(\dot{m}^{\rm r}|L\bol)$ at the given bolometric luminosity
$L\bol=10^{43.75}$ (solid line), $10^{44.25}$ (dotted line), $10^{44.75}$
(short-dashed line), $10^{45.25}$ (long-dashed line), $10^{45.75}$
(dot-short-dashed line), $10^{46.25}$ (dot-long-dashed line), $10^{46.75}$
(short-dash-long-dashed line), and $10^{47.25}\ergs$ (solid line),
respectively. See details in \S~\ref{sec:accrete}.
}                                                             
\label{fig:pmdotLbol}                                         
\end{center}                                                  
\end{figure}

Adopting the reference model for the luminosity evolution of individual QSOs,
we calculate the probability distribution of underlying Eddington ratios among
QSOs at a given $L\bol$. As shown in Figure~\ref{fig:pmdotLbol}, the
probability of finding objects with low Eddington ratios in low-luminosity QSOs
is larger than that in high-luminosity QSOs. The average Eddington ratio in
high-luminosity QSOs is larger than that in low-luminosity QSOs and the width
of the Eddington-ratio distribution in high-luminosity QSOs is narrower than
that in low luminosity QSOs, although the Eddington-ratio (or the
accretion-rate) evolution in individual QSOs is assumed to be uniform.  The
$\delta$-function like distribution at $\dot{m}^{\rm r}=1$ [where we use
$\frac{1}{a\sqrt{\pi}} \exp(-x^2/a^2)$ with $a=0.1$ to mimic the Dirac function
$\delta(\dot{m}^{\rm r}-1)$ for convenience] for each given $L\bol$ represents
the self-regulated rapid accretion phase with a rate close to the Eddington
limit when the accretion material is over-supplied. Given an $L\bol$, a lower
Eddington ratio corresponds to a higher MBH mass, and the exponential decline
of the probability distribution of Eddington ratios at small $\dot{m}^{\rm r}$
for QSOs at a given $L\bol$ is primarily due to the exponential-like decay of
MBH abundance at the high-mass end ($\bh > 10^8\msun$). The Eddington ratios in
most of the luminous QSOs ($L\bol\ga 10^{45.75}\ergs$) are close to $1$ because
of the steep falloff of the BHMF at the high-mass end and the rapid decay of
Eddington ratios with time in the declining phase of the accretion-rate
evolution in individual QSOs.  These underlying Eddington-ratio distributions
are clearly different from those observational estimates, i.e., a Gaussian
distribution of $\dot{m}^{\rm obs}$ with peaks around $10^{-0.6}-10^{-1.1}$
\citep{Kollmeier06, Shen07, Netzer07}.

The observationally estimated Eddington-ratio distribution may be biased from
the underlying true Eddington ratio distribution. The reasons are: (1) the
masses of MBHs in QSOs are usually obtained by using the virial mass
estimator(s) $\bh^{\rm vir}$, and the virial mass estimator is based on the
analysis of broad emission line reverberation mapping data for several tens of
low-luminosity AGNs at low redshift and a calibration of it to the local
$\mbh-\sigma$ relation.  The estimates of $\bh^{\rm vir}$ may scatter around
and be offset from the real $\bh^{\rm r}$, as the relation between luminosity
and broad line region (BLR) size and the relation between FWHM of emission
lines and BLR virial velocity, adopted in the virial mass estimator(s), are not
perfect, and its validation for high luminosity QSOs at high redshift is not
fully tested \citep[e.g.,][]{Kaspi07}.  A scatter of $0.3$~dex in inferred
$\bh^{\rm vir}$ is plausible as pointed out by \citet{Kollmeier06} (see also
\citealt{Shen07}) because the relation between observed line width and MBH mass
may depend on the viewing angle of BLR \citep[e.g.,][]{Krolik01} and the
relation between BLR size and luminosity has an intrinsic scatter about
$0.1-0.2$~dex \citep{Kaspi05}. (2) There may be some systematic errors as large
as a factor of 3 or more either up or down in the virial mass estimator(s) due
to various effects, such as, a broad radial emissivity distribution, and an
unknown angular radiation pattern of line emission \citep[see][]{Krolik01}.
These systematic errors may introduce an offset of the virial mass estimator(s)
from the underlying true mass.  (3) The bolometric luminosities are usually
obtained using a uniform bolometric correction \citep[see][]{Kollmeier06,
Shen07}, but the real bolometric corrections may scatter around this uniform
mean value by $\sim 0.1$~dex in the optical band (see \S~\ref{subsec:Bolcor}).
The dominant bias is probably those introduced by the viral mass estimator(s).

We assume that the probability distribution of $\bh^{\rm vir}$ for a given
underlying real MBH mass $\bh^{\rm r}$ is
\begin{eqnarray}
& & P(\log\bh^{\rm vir}|\log\bh^{\rm r})  = \frac{1}{\sqrt{2\pi}
\Delta_{\log\bh^{\rm vir}}}\times \nonumber \\
& & \exp\left[-\frac{(\log\bh^{\rm vir}-\log\bh^{\rm r}-
\Theta_{\log\bh^{\rm vir}})^2}{2\Delta^2_{\log\bh^{\rm vir}}}\right],
\label{eq:revbscat}
\end{eqnarray}
where $\Delta_{\log\bh^{\rm vir}}$ is the scatter of MBH masses estimated by
using the virial mass estimator(s) around the underlying given true mass, and
$\Theta_{\log\bh^{\rm vir}}$ is the offset of $\log\bh^{\rm vir}$ from the true
mass of MBHs $\log\bh^{\rm r}$. As discussed above, it is plausible that
$\Delta_{\log\bh^{\rm vir}}\sim 0.3$~dex and $\left | \Theta_{ \log\bh^{\rm
vir}}\right | \la (0.3-0.6)$ \citep[e.g.,][]{Krolik01,Kollmeier06,Shen07}.  For
a given $\dot{m}^{\rm r}$ at fixed $L\bol$ and $\mbh$, the observationally
estimated Eddington ratio is thus given by
\begin{eqnarray}
& & P(\dot{m}^{\rm obs}| \dot{m}^{\rm r})=
\frac{1}{\sqrt{2\pi}\ln(10)\Delta_{\log\bh^{\rm vir}}\dot{m}^{\rm
obs}} \times \nonumber \\
 & & \exp\left[-\frac{(\log \dot{m}^{\rm obs}-\log\dot{m}^{\rm
r}+\Theta_{\log\bh^{\rm vir}})^2}{2 \Delta^2_{\log\bh^{\rm vir}}}\right],
\label{eq:mdotobsmdot}
\end{eqnarray}
Combining this probability distribution with equation (\ref{eq:probmdotr}),
the `time-integrated' observational Eddington ratio ($\dot{m}^{\rm obs}$)
distribution can be inferred from the local BHMF as
\begin{eqnarray}
& & P(\dot{m}^{\rm obs}|L\bol)=\int d\dot{m}^{\rm r} P(\dot{m}^{\rm 
obs}|\dot{m}^{\rm r})\int  n(\mbh) \times \nonumber \\
& & \tau\life(\mbh)P(L\bol|\mbh) P(\dot{m}^{\rm r}|L\bol,\mbh)d\mbh \times 
    \nonumber \\
& & \left[ \int n(\mbh)\tau\life(\mbh)P(L\bol|\mbh) d\mbh\right]^{-1},
\label{eq:probmdotobs}
\end{eqnarray}
provided that the accretion rate or luminosity evolution of individual QSOs,
i.e., ${\cal L}(\tau;\mbh)$ and thus $P(L\bol|\mbh)$, is known.

\begin{figure}
\begin{center}
\includegraphics[width=\hsize,angle=0]{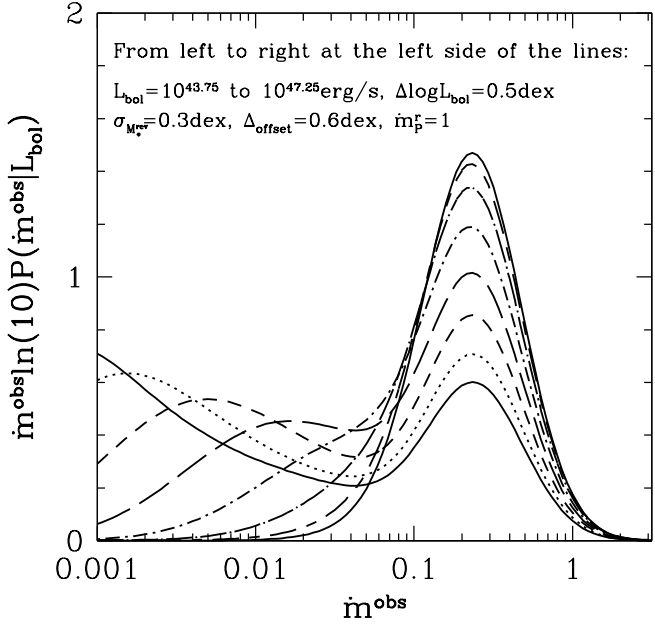}
\caption{The observational `time-integrated' Eddington ratio distribution
among QSOs at a given bolometric luminosity inferred from the reference
model in \S~\ref{sec:models} by assuming that the scatter and offset
in the masses of MBHs obtained from the virial mass estimator(s) are
$0.3$~dex and $0.6$~dex, respectively. The line types have the same meaning
as in Fig.~\ref{fig:pmdotLbol}.
}
\label{fig:pmdotobs}
\end{center}
\end{figure}

We show $P(\dot{m}^{\rm obs}|L\bol)$ calculated from the reference model in
Figure~\ref{fig:pmdotobs}, using equation~(\ref{eq:probmdotobs}) and assuming
$\Delta_{\log\bh^{\rm vir}}=0.3$~dex and $\Theta_{\log\bh^{\rm vir}}=0.6$~dex
(for which the Eddington ratio in the first rapid accretion phase is
$\dot{m}^{\rm r}=\dot{m}^{\rm r}_{P}=1$). It appears that the
`time-integrated'  Eddington-ratio distribution is approximately a Gaussian
distribution at any fixed bolometric luminosity $L\bol\ga 10^{45.75}\ergs$ but
with a small tail at the low-Eddington ratio end. The Gaussian-like
distribution mainly corresponds to the peaks at $\dot{m}^{\rm r}=1$ as shown in
Figure~\ref{fig:pmdotLbol}, which represent the rapid accretion phase of
individual QSOs with a rate self-regulated by the Eddington limit. The width of
the Gaussian-like distribution mainly reflects the scatter
$\Delta_{\log\bh^{\rm vir}}$ in the estimates of MBH masses using the virial
mass estimator(s), and the locations of peaks in the Eddington ratio
distribution are roughly determined by the offset $\Theta_{\log\bh^{\rm vir}}$
and the value of Eddington ratio $\dot{m}^{\rm r}_{P}$ during the
self-regulated rapid accretion phase when the accretion material is
over-supplied. For QSOs with lower bolometric luminosities ($L\bol\la
10^{45.25}\ergs$), the probability of finding low Eddington-ratio
($\dot{m}^{\rm obs}\la 0.03$) objects becomes significant, which is primarily
because the underlying MBH mass function is shallow at the low-mass end
($\bh\la 10^8\msun$) and the decline phase of the self-similar evolution of the
disk accretion (see also Fig.~\ref{fig:lc}) around big MBHs contributes
significantly to the counts of low bolometric luminosity objects. Therefore,
the Eddington ratio distribution among low-luminosity QSOs should provide
independent constraints on the long-term evolution of disk accretion,
especially in the decline phase (see also \citealt{YLK05}).

Except for giving a rough comparison with observations below, we do not intend
to use $P(\dot{m}^{\rm obs}|L\bol)$ inferred from the reference model in this
paper to directly fit the observational Eddington ratio distribution estimated
by \citet{Kollmeier06}, \citet{Shen07}, and \citet{Netzer07}. The reasons are:
(1) $P(\dot{m}^{\rm obs}|L\bol)$ obtained from equation (\ref{eq:probmdotobs})
is a `time-integrated' and volume-weighted Eddington-ratio distribution, while
the current observationally estimated distributions are for given redshift
intervals and not volume weighted. If the Eddington-ratio distribution at a
given bolometric luminosity is independent of redshift as suggested by
\citet{Kollmeier06} (but perhaps it is not as argued by \citealt{Shen07}), the
`time-integrated' Eddington ratio distribution among QSOs may represent the
observational one for different redshift intervals; (2) the observational
Eddington ratio distribution may be biased significantly at low Eddington
ratios in flux-limited surveys, such as the SDSS (see \citealt{Kollmeier06} and
\citealt{Shen07}; for more general discussion of selection bias see also
\citealt{Lauer07b} and \citealt{YL04b}); (3) $P(\dot{m}^{\rm obs}|L\bol)$
obtained from equation (\ref{eq:probmdotobs}) does not distinguish obscured and
unobscured QSOs,  while the observationally estimated distribution is primarily
obtained from optical QSO samples. If the obscuration is only a geometrical
effect, then $P(\dot{m}^{\rm obs}|L\bol)$ may be the same as the
Eddington-ratio distribution in optical QSO samples; however, if the
obscuration is partly due to an evolutionary effect (e.g., QSOs may be more
likely obscured in their early accretion stage), the probability of QSOs with
high Eddington ratios may be suppressed.

If the MBH masses inferred from the virial mass estimator have an offset $\sim
0.5-0.6$~dex from the underlying true masses and also a scatter of $\sim
0.3$~dex around them, $P(\dot{m}^{\rm obs}|L\bol)$ at any fixed bolometric
luminosities $L\bol\ga 10^{45.75}\ergs$ is roughly consistent with the
Gaussian-like distribution of Eddington ratios (with a scatter of $\sim
0.3$~dex, and peak at $1/4$) among QSOs estimated by \citet{Kollmeier06};
however, $P(\dot{m}^{\rm obs}|L\bol)$ at bolometric luminosity $L\bol\la
10^{45.25}\ergs$ has a tail extended to lower Eddington ratios ($\dot{m}^{\rm
obs}\la 0.01$), which appears to conflict with that obtained by
\citet{Kollmeier06} for QSOs with $L\bol<10^{45.5}\ergs$.  The Eddington ratio
in the self-regulated rapid accretion phase is assumed to be $1$ (i.e., exactly
the Eddington limit) in the above calculations, but it may be slightly
different from $1$ and a little smaller. For example, if $\dot{m}^{\rm
r}=\dot{m}^{\rm r}_{P}=0.5$ in the first rapid accretion phase of the model
(c), the time-integral of XAGN LF inferred from the local BHMF can still
marginally match that obtained from observations\footnote{Our calculations
show that this set of the accretion rate in the first rapid accretion phase
seems to under-predict the time-integral of the optical QSO LF at the
high-luminosity end ($M_{\rm B}\la -28$), which may be partly due to some
uncertainties in the bolometric correction for the $B$ band at the
high-luminosity end; otherwise QSOs must accrete at a rate closer to the
Eddington limit during the first rapid accretion phase.}, but it cannot if
$\dot{m}^{\rm r}=\dot{m}^{\rm r}_{P}$ is substantially less than $0.5$.
If $\dot{m}^{\rm r}_{P}=0.5$, to reproduce the observational Eddington
ratio distribution at high bolometric luminosity ($L\bol\ga 10^{45.75}\ergs$),
$\Theta\sim 0.2-0.3$~dex is required; and the tail of $P(\dot{m}^{\rm
obs}|L\bol)$ at low Eddington ratios for low-luminosity QSOs ($L\bol\la
10^{45.25}\ergs$) is still significant, though less significant compared
to the case for $\dot{m}^{\rm r}=\dot{m}^{\rm r}_{P}=1$ in the first rapid
accretion phase.

The consistency between $P(\dot{m}^{\rm obs}|L\bol)$ inferred from the extended
So{\l}tan argument and the observationally estimated Eddington-ratio
distribution at high bolometric luminosity ($L\bol\ga 10^{45.75}\ergs$)
suggests that the majority of bright QSOs accrete material at a single rate
close to the Eddington limit.  But we should be cautious of any
over-interpretation of the possible inconsistency at low bolometric luminosity
($L\bol\la 10^{45.25}\ergs$) above, since the observational results obtained by
different authors have not yet converged \citep[e.g.,][]{Kollmeier06, Shen07,
Netzer07}.  For example, the observational Eddington-ratio distribution among
QSOs with $L\bol<10^{45.5}\ergs$ obtained by \citet{Shen07} (i.e., a
Gaussian-like distribution with peak at $10^{-1.1}$ and a scatter of
$0.42$~dex) is substantially shifted to lower Eddington ratios compared with
that obtained by \citet{Kollmeier06} (i.e., a Gaussian-like distribution with
peak $10^{-0.6}$ and scatter $0.3$~dex).  Note also that the Eddington ratios
in low-luminosity AGNs at low redshifts do cover a wide range as shown by
\citet{WU02}, \citet{Heckman04}, and \citet{GreeneHo07}, which may be
consistent with the prediction obtained from equation (\ref{eq:probmdotobs})
above.

The offset $\Theta_{\log\bh^{\rm vir}}\sim 0.3-0.6$~dex required by the above
observational constraints suggests that the MBH masses inferred from the virial
mass estimator(s) may be over-estimated by a factor of $2-4$, at least for
high-luminosity QSOs, which is compatible with the possible systematic errors
in the MBH mass estimated by the reverberation mapping technique
\citep{Krolik01}. If this offset is real, it is intriguing and important since
many current studies on the growth and evolution of MBHs in QSOs are based on
the virial mass estimator(s). For example,  the masses of MBHs in two samples
of AGNs at redshifts $z=0.36$ and $z=0.57$, estimated from the virial mass
estimator(s), are found to be larger than that estimated from the local
$\mbh-\sigma$ relation by $0.54$~dex and $0.51$~dex, respectively
\citep{Treu04,Woo06,Woo08}, which is suggested as an indicator of that the
growth of MBHs predates the final growth of bulges in these AGN host galaxies.
If MBH masses from the virial mass estimator(s) are generally over-estimated by
a factor of $2-4$ as argued above, then there should be not much difference
between the rescaled virial masses and that predicted from the $\bh-\sigma$
relation for those MBHs in the studies of \citet{Treu04} and
\citet{Woo06,Woo08}.

Similar to the probability distribution of the Eddington ratio among QSOs at
a given bolometric luminosity given in equation (\ref{eq:probmdotr}), the
probability distribution of the Eddington ratio among QSOs at a given MBH mass
$\bh^{\rm r}$ can be estimated by
\begin{eqnarray}
& & P(\dot{m}^{\rm obs}|\bh^{\rm r})=\int d\dot{m}^{\rm r}
P(\dot{m}^{\rm obs}|\dot{m}^{\rm r})\int  n(\mbh)\times \nonumber \\
& & \tau\life(\mbh)P(\bh^{\rm r}|\mbh) 
P(\dot{m}^{\rm r}|\bh^{\rm r},\mbh)d\mbh \times\nonumber \\
& & \left[\int n(\mbh)\tau\life(\mbh)P(\bh^{\rm r}|\mbh) d\mbh\right]^{-1},
\label{eq:probmdotmr}
\end{eqnarray}
or the probability distribution of Eddington ratio among QSOs at a given
$\bh^{\rm vir}$ can be estimated as
\begin{eqnarray}
P(\dot{m}^{\rm obs}|\bh^{\rm vir})=\int d\dot{m}^{\rm r}
P(\dot{m}^{\rm obs}|\dot{m}^{\rm r}) \int d\bh^{\rm r}
P(\bh^{\rm vir}|\bh^{\rm r}) \times & & \nonumber  \\
\int  n(\mbh)\tau\life(\mbh) P(\bh^{\rm r}|\mbh) 
P(\dot{m}^{\rm r}|\bh^{\rm r},\mbh)d\mbh / & & \nonumber \\
\int P(\bh^{\rm vir}|\bh^{\rm r})d\bh^{\rm r} \int
n(\mbh)\tau\life(\mbh)P(\bh^{\rm r}|\mbh) d\mbh, & & \nonumber \\
\label{eq:probmdotmobs}
\end{eqnarray}
where $P(\bh^{\rm vir}|\bh^{\rm r})=P(\log\bh^{\rm vir}|\log\bh^{\rm r})/
[\bh^{\rm vir}\ln(10)]$.  We note here that $P(\dot{m}^{\rm obs}|\bh^{\rm
vir})$ should be skewed toward low Eddington ratios because a QSO may spend
a majority of its lifetime in the declining phase with small $\dot{m}^{\rm r}$
($<0.1$) as that revealed by the reference model in this paper (e.g.,
Fig~\ref{fig:lc}). However, it is not easy to observationally estimate
$P(\dot{m}^{\rm obs}|\bh^{\rm vir})$ since low-luminosity QSOs are more likely
to be missed in flux-limited surveys, especially at high redshift
\citep[see also discussions in][]{Kollmeier06}. We defer the
comparison of this distribution with observations to future work.

\section{Toy models for mergers} \label{sec:BHmerger}

\subsection{Gas-poor (dry) mergers}

In current hierarchical galaxy formation models, mergers of galaxies are the
main route to form elliptical galaxies and stellar bulges
\citep[e.g.,][]{KWG93,Cole00}. If each merging galaxy has a central MBH,
mergers of two galaxies will inevitably form binary MBHs and may further lead
to mergers of MBHs if their inspiral and orbital evolution time is shorter than
a Hubble time \citep[e.g.,][]{BBR80,Y02}.  Mergers of MBHs occurred after the
quenching of nuclear activity, as the probable consequence of galaxy gas-poor
(dry) mergers, may significantly re-shape the BHMF established by accretion
processes. Below we use a toy model to illustrate the change of the BHMF due to
BH mergers under the assumption that MBHs grow only by merging BHs but with
little accretion directly onto MBHs during the galaxy dry merger stage.

Assume that a major dry merger of two host galaxies always leads to the
merger of their central MBHs with mass ratio of $\alpha=\bhs/(\bhp+\bhs)$
(e.g., 0.5 or 0.25 for a 1:1 or 1:3 merger) after the nuclear activity
is quenched, where $\bhp$ and $\bhs$ are the masses of the two BHs before their
merger. The merged BH mass is $\mbh=\beta(\bhp+\bhs)$ and ($1-\beta$) is the
fraction of energy (or mass) losses due to gravitational waves during the MBH
merger process.  Thus, the mass function of QSO remnants right after the
nuclear activity is 
\begin{eqnarray}
n'_{\bh}(\mbh')&=&\int d\mbh n_{\bh}(\mbh) \times \nonumber \\
& & \left[\delta(\mbh'-\frac{\alpha}{\beta}\mbh)+
  \delta(\mbh'-\frac{1-\alpha}{\beta}\mbh) \right] \nonumber\\
&=&\frac{\beta}{\alpha}n_{\bh}(\mbh) 
   |_{\mbh=\frac{\beta}{\alpha}\mbh'}+ \nonumber \\
& &    \frac{\beta}{1-\alpha}
   n_{\bh}(\mbh)|_{\mbh=\frac{\beta}{1-\alpha}\mbh'}.
\label{eq:BHMFmrg}
\end{eqnarray}
Note this mass function $n'_{\bh}(\mbh')$ is non-synchronous since the
last major (dry) merger may occur at different time for different MBHs.

During the merging process of BHs (which is divided into three phases:
inspiral, merger and ringdown), the total energy lost through gravitational
waves, $E_{\rm rad}$ is difficult to calculate, especially for the merger of
two BHs with large spins, but roughly in the range $0.03M_{\bullet,12}
F(\mu/M_{\bullet,12})<E_{\rm rad}<0.2M_{\bullet,12}F(\mu/M_{\bullet,12})$,
where $M_{\bullet,12}= \bhp+\bhs$ is the total (initial) mass of the two BHs,
$\mu$ is the reduced mass and $F(\mu/M_{\bullet,12})=(4\mu/M_{\bullet,12})^2$
\citep[see eq.~3.7 in][]{FH98}.  Recent breakthrough in relativistic numerical
calculation of merging binary BHs due to \citet{Pretorius05} and
\citet{Baker06} has shown that on the order of 5\% of the initial rest mass for
a system of two equal mass, nonspinning BHs is radiated as gravitational waves
during the final orbit and ringdown, which is consistent with the estimate in
\citet{FH98}. For the merging of equal mass, rapidly spinning BHs,  the energy
radiated as gravitational wave could be larger and the upper limit is about
24\% if the final spin is around $0.9$ \citep[e.g.,][]{PK07}. Since the spin of
MBHs is probably close to $0.7-0.9$ due to accretion processes
\citep[e.g.,][]{Gammie04, Shapiro05, HBK07}, here we choose two cases, 10\% and 24\%
of the initial total rest mass, for the amount of energy radiated as
gravitational waves, and therefore $\beta=1-0.1F(\mu/M_{\bullet,12})$ and
$1-0.24F(\mu/M_{\bullet,12})$, respectively.

\begin{figure}
\begin{center}
\includegraphics[width=\hsize,angle=0]{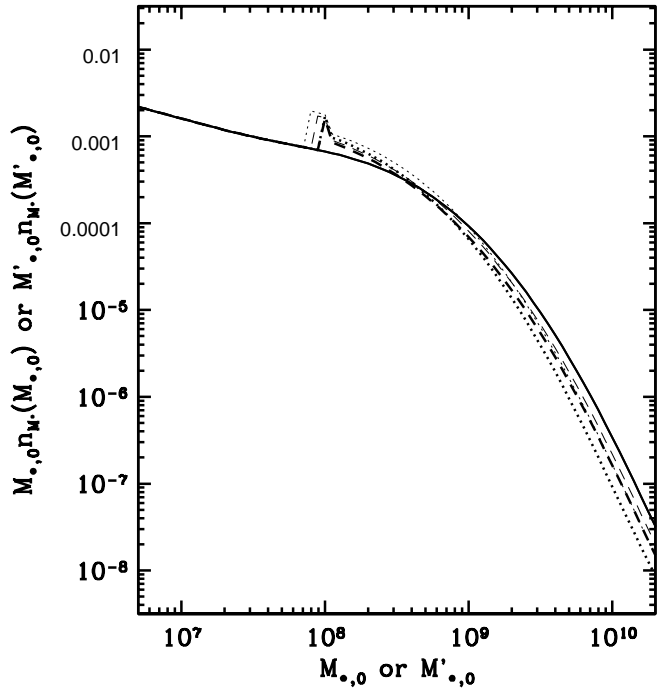}
\caption{
Effect of dry mergers on re-shaping the BHMF after the quenching of nuclear
activity. The solid line represent the BHMF $\mbh n_{\bh}(\mbh)$ in the local
universe estimated by the $\mbh-\sigma$ relation and the velocity-dispersion
distribution as that shown in Fig.~\ref{fig:lmf}. The dashed and dotted lines
show the BHMF $n_{\bh}(\bh')$ after correcting the effect due to dry mergers.
Dotted (or dashed) lines represent the case that all MBHs with mass
$>10^8\msun$ experienced one 1:1 (or 1:3) dry major merger, while other MBHs
with smaller mass have not undergone any major mergers, after the quenching of
nuclear activity.  The loss of energy through gravitational waves are assumed
to be 10\% (thick dotted or dashed lines) or 24\% (thin dotted or dashed lines)
of the initial total rest mass.  This figure shows that the abundance of MBHs
after the quenching of nuclear activity may be smaller than that estimated in
the local universe at most by a factor of 2--3 at the high-mass end, but is not
significantly different from the local BHMF at the low-mass end. This
difference caused by the effect of dry mergers on the estimated BHMF is not so
prominent compared with that due to the uncertainty of intrinsic scatters in
the $\mbh-\sigma$ and $\mbh-L\bulge$ relations. 
}
\label{fig:BHMFmrg}
\end{center}
\end{figure}

Using the data from Galaxy Evolution from Morphology and SEDs (GEMS),
\citet{Bell06} find that present-day spheroidal galaxies with $M_V<-20.5$ on
average have undergone between 0.5 and 2 major dry mergers since redshift
$z<0.7$ (see also similar results in \citealt{Conselice03} for redshift $z\la
3$, and \citealt{Lin04} for redshift $z\la 1.2$). MBHs with mass substantially
less than $10^8\msun$ are mostly hosted by the stellar bulges of spiral
galaxies, which should not have undergone a significant number of major dry
mergers in the near past (e.g., $z\la 1$) since their disks are preserved.  For
simplicity, here we assume all MBHs with $\mbh\ga 10^8\msun$, corresponding to
$M_V<-20.5$, have experienced one major dry merger after the quenching of
nuclear activity, while smaller MBHs did not experience major dry mergers.
Using equation (\ref{eq:BHMFmrg}) to correct the effect due to dry mergers in
$n_{\bh}(\mbh)$, the BHMF right after the quenching of nuclear activities,
$n_{\bh}(\mbh')$, which is established by the accretion process, is estimated
and shown in Figure~\ref{fig:BHMFmrg}.  As seen from Figure~\ref{fig:BHMFmrg},
the abundance of MBHs with mass larger than a few $10^9\msun$ may be enhanced
at most by a factor $\sim 2-3$ due to the major dry mergers after the quenching
of nuclear activities. This enhancement is not so prominent compared with that
in the estimate of BHMF due to the uncertainty in the intrinsic scatters in the
$\mbh-\sigma$ and $\mbh-L\bulge$ relationships.  For example, a slight error in
the estimated intrinsic scatters, e.g., $0.05$~dex, could introduce an
uncertainty larger than a factor of $\sim 2-3$ at the high-mass end of the
BHMF. Therefore, we conclude that it is safe to neglect the uncertainty in the
BHMF due to major dry mergers after the quenching of nuclear activities in our
calculations as the intrinsic scatters in the $\mbh-\sigma$ or
$\mbh-L\bulge$ relations are currently not well determined.

\subsection{Gas-rich (wet) mergers}

For a present-day MBH with mass $\mbh$, its host galaxy may have undergone more
than one (wet) major merger. By `wet' major merger here we mean gas-rich major
mergers (spiral+spiral or spiral+elliptical) which lead to substantial gas
fueling to trigger nuclear activity. In the current scenario of hierarchical
formation and co-evolution of galaxies and MBHs \citep[e.g.,][]{KH00, Bower06,
Croton06, Malbon07}, each `wet' major merger leads to a nuclear active phase
with substantial increase in the central MBH mass.  For a local MBH we see
today, it may build up through multiple times of nuclear activity and mergers
of smaller MBHs, and thus our assumption of a single time of nuclear activity
in \S~\ref{sec:models} may be an over-simplification. In order to estimate the
effects of possible multiple phases of nuclear activity in the assembly history
of a MBH, we assume that all MBHs experienced two or more `wet' major mergers
and its mass increases by a factor of $5$, $10$ or more after each `wet' major
merger, and this assumption is compatible with the current co-evolution models
\citep[e.g.][]{KH00, Bower06, Croton06, Malbon07}. Before each `wet' major
merger, we assume that the MBH triggered by the `wet' merger later is the
merger remnant of two progenitors with mass ratio either $1:1$ or $1:3$.  These
two progenitors also experienced a similar period of significant mass growth
before and so on and so forth. We can use this procedure backwards for two or
more times to mimic multiple phases of accretion (including mergers). With the
above assumptions, combining equations (\ref{eq:BHMFmrg}) and
(\ref{eq:intYband}), the time-integral of XAGN LF can be calculated. In these
calculations we also adopt the same parameters ($\epsilon$, $\zeta$, $\gamma$)
as in the reference model but adjust $\xi$ in order to satisfy the assumption
of a factor of $5$, $10$ or more mass increase during each `wet' major merger.
We find that the inferred value of the time-integral of XAGN LF at the
low-luminosity end for the multiple-times nuclear activity assumption is larger
than that obtained by assuming a single-time nuclear activity at most by $\sim
20-30\%$, and the difference is negligible at the high-luminosity end.  We
therefore conclude that the assumption of a single-time nuclear activity is
good enough, provided that the masses of MBHs substantially increase (say, by a
factor of $5$, $10$ or more) during the nuclear activity.

\section{Further Implications}\label{sec:discon}

\subsection{BHMF at redshift $z$}

If the probability function $P(L|\mbh)$ is known (for example, as described by
the reference model we obtained from the extended So{\l}tan argument above),
the BHMF at redshift $z$, i.e., $n_{\bh}(\mbh,t_z)$ can be estimated by an
equation similar to equation~(\ref{eq:relation})
\begin{eqnarray}
\int^{\infty}_z \Psi_{L}(L,z)\left|\frac{dt}{dz}\right|dz&=&\int^{\infty}_{0} 
n_{\bh}(\mbh,t_z)\tau\life(\mbh) \times \nonumber \\
& & P(L|\mbh) d\mbh,
\label{eq:relationz}
\end{eqnarray}
where $t_z=\int^{\infty}_{z}\left|\frac{dt}{dz}\right|dz$ is the cosmic time at redshift $z$. Equation (\ref{eq:relationz})
is true only if the BHMF at $z$ is dominated by quiescent MBHs.

If MBHs with final mass $\mbh$ shined for a time $\tau\life(\mbh)$ with
luminosity $L=\lambda L\Edd(\bh)$ and without increasing their mass
significantly ($\mbh\simeq\bh$), where $\lambda$ is a constant and $L\Edd(\bh)$
is the Eddington luminosity (see definition in eq.~\ref{eq:Ledd}), then
$P(L|\mbh)\simeq \delta[L-\lambda L\Edd(\mbh)]$ and thus
\begin{eqnarray}
n_{\bh}(\mbh,t_0) &\simeq& \frac{1}{\tau\life(\mbh)}
\int^{\infty}_0 \Psi_{L}(L,z)|_{L=\lambda L_{\rm Edd}(\mbh)}\times
\nonumber \\
& &
\left|\frac{dt}{dz}\right|dz
\times \lambda \left|\frac{dL_{\rm Edd}(\bh)}{d\bh}\right|_{\bh=\mbh}.
\label{eq:nmt0}
\end{eqnarray}
If these assumptions are correct, it appears that the local BHMF can be forced
to be always consistent with the QSO LF by adjusting $\tau\life(\mbh)$ and
$\lambda$.  Similarly, the BHMF at redshift $z$ is given by
\begin{eqnarray}
n_{\bh}(\mbh,t_z) &\simeq& \frac{1}{\tau\life(\mbh)}
\int^{\infty}_z \Psi_{L}(L,z)|_{L=\lambda L_{\rm Edd}(\mbh)} \times \nonumber \\
& &
\left|\frac{dt}{dz'}\right|dz'
\times \lambda \left|\frac{dL_{\rm Edd}(\bh)}{d\bh}\right|_{\bh=\mbh}.
\label{eq:nmtz}
\end{eqnarray}
The BHMF at redshift $z$, $n_{\bh}(\mbh,t_z)$, estimated from
equation (\ref{eq:nmtz}) may be not accurate if the real $\tau\life(\mbh)$ is
substantially longer than the Salpeter timescale since in this case the MBH
mass was evolving rapidly during its active phase and the BHMF sharply
decreases at the high-mass end, and also the assumption of $P\simeq \delta(L-
\lambda L\Edd)$ is not good. The errors in the BHMF at redshift $z$
obtained from this simple approach can be estimated by comparing it with that
obtained from equation (\ref{eq:relationz}).  With the constraints on the
luminosity evolution of individual QSOs obtained above, we will estimate the
BHMF at different redshift $z$ and check whether these estimates are consistent
with observations in a future study.

\subsection{Triggering rate of nuclear activity}

The QSO LF at redshift $z$ can be inferred as
\begin{eqnarray}
\Psi_L(L,z) &=& \int d\mbh \int^{\infty}_{z} 
{\cal G}(z_i;\mbh) \delta(L-{\cal L}(\tau;\mbh))\times \nonumber \\
& & \left|\frac{dt}{dz_i}\right|dz_i 
\label{eq:QSOLFz}
\end{eqnarray}
where $\tau=t_z-t_{z_i}$. Given $\Psi_L(L,z)$ and ${\cal L}(\tau;\mbh)$, the
triggering rate of nuclear activity in the mass range $\mbh-\mbh+d\mbh$,
${\cal G}(z;\mbh)$, can be solved from the above integral equation.  Estimation of
${\cal G}(z;\mbh)$ is of fundamental importance because it is this function that
dominates the cosmic evolution of the QSO population and the down-sizing nature
of the formation of MBHs. Here the cosmic evolution of the QSO population means
that the comoving number density of the QSO population brighter than a certain
luminosity (or in a certain luminosity range) has a peak at an intermediate
redshift (e.g., $z\sim 2-3$) and decreases at both higher and
lower redshift \citep[e.g.,][]{Richards06a,Hasinger05} and the down-sizing nature refers to that observationally the
characteristic mass of MBHs in QSOs decreases with decreasing redshifts
\citep[e.g.,][]{Marconi04,Merloni04}. Given
${\cal G}(z;\mbh)$ and ${\cal L}(\tau;\mbh)$, many statistical properties of QSOs (for
instance, the Eddington rate ratio distribution and the MBH mass distribution
at different redshifts in QSOs/AGNs) can be inferred.  Comparison of these
inferred properties with those directly obtained from observations will further
reveal details of the growth and evolution of MBHs and QSOs. We will present
this in a future study.

\section{Conclusions}\label{sec:conclusion}

In this paper, we have studied the observational constraints on the growth of
MBHs using the extended So{\l}tan argument. In this approach, the local BHMF is
directly connected with the time-integral of the QSO LF through only the
luminosity evolution of individual QSOs (and correspondingly the accretion-rate
evolution, given the mass-to-energy conversion efficiency), and the luminosity
evolution of individual QSOs is isolated from the cosmic evolution of the
triggering rate of nuclear activity.  The luminosity (or accretion-rate)
evolution of individual QSOs has an unambiguous physical definition, and it is
different from the `mean accretion rate' as a function of mass and/or redshift
widely used in the literature \citep[e.g.,][]{Marconi04, Shankar04, Shankar07}
in that the `mean accretion rate' is a combined property depending on both the
luminosity evolution of individual QSOs and the cosmic evolution of the
triggering rate of nuclear activity.

With recent knowledge of the relationships between MBH mass and host galaxy
properties (either the $\mbh-\sigma$ relation or the $\mbh-L_{\rm bulge}$
relation) and the distribution of galaxy properties (either $\sigma$ or $L_{\rm
bulge}$), we estimate the local BHMF. We obtain the time-integral of the QSO LF
from recent estimates of QSO LFs in both optical and X-ray bands. Using the
local BHMF and the time-integral of the QSO LF, we obtain robust constraints on
the luminosity (or accretion rate) evolution of individual QSOs and important
characteristic parameters describing the growth of individual MBHs, such as
the mass-to-energy conversion efficiency and lifetime, which are summarized
below.

\begin{itemize}

\item The luminosity (or accretion rate) evolution of individual QSOs probably
involves two phases: an initially exponentially increasing phase set by the
Eddington limit (i.e., ${\cal L}\simeq L\Edd$) when the infall material to feed
the central MBHs is over-supplied; and then followed by a phase with power-law
declining  set by a self-similar long-term evolution of disk accretion (i.e.,
${\cal L}\propto \tau^{-\gamma}$ and $\gamma\sim 1.2-1.3$).  With this type of
luminosity evolution, the time-integral of QSO LF can be well matched by that
inferred from the local BHMF. Other simple luminosity evolution models, such as
a single Eddington ratio for all MBHs/QSOs or an initially exponentially
increasing phase followed by an exponentially decay phase, cannot satisfy the
extended So{\l}tan argument simultaneously at both the high-luminosity end and
low-luminosity ends, and thus are ruled out.

\item The mass-to-energy conversion efficiency $\epsilon$ is $\simeq 0.16\pm
0.04^{+0.05}_{-0}$ (correspondingly the  spin parameter $a$ is in the range from $0.8$ to
$0.99$) if adopting the local BHMF estimated from the $\mbh-\sigma$
relation, which is fully consistent with the theoretical expectations of $\sim
0.10-0.20$, i.e., the spin of MBHs in QSOs may stay at an equilibrium of $\sim
0.7-0.9$ for most of the QSO lifetime \citep[e.g.,][]{Gammie04, Shapiro05, HBK07}.
However, the efficiency $\epsilon$ is reduced to $\simeq 0.08 \pm
0.02^{+0.03}_{-0}$ (and correspondingly the spin parameter $a$ is in the range from $0.1$ to
$0.8$) if adopting the local BHMF estimated from the $\mbh-L_{\rm
bulge}$ relation, which is lower than but may be still marginally consistent
with theoretical expectations.

\item The lifetime of QSOs/AGNs, which depends on detailed definition of the
nuclear activity or the lower threshold set to the active nuclear luminosity,
can be as long as a few $10^9$~yr, and the characteristic timescale in
the luminosity increasing phase and transition timescale to the declining phase
do not necessarily depend on the mass of their central MBHs. The period that a
QSO or MBH radiating at a luminosity larger than $10\%$ of its peak luminosity
is only about $2-3\times 10^8$~yr, and during this period the MBH obtained most
of its mass. If adopting the local BHMF estimated from the $\mbh-L_{\rm bulge}$
relation, the above values related to the QSO lifetime decrease by a factor
$\sim 2$.

\item For individual QSOs, the characteristic timescale for the luminosity 
(or accretion rate) to decline from its peak $L_{P}$ to $0.1L_{P}$ in
their second phase should be relatively short compared to the Salpeter
timescale, which suggests that the material infalling from a large galactic
scale and deposited in the vicinity of MBHs be consumed by rapid accretion onto
the central MBH and at the mean time further deposit of material can be
efficiently suppressed by some mechanisms, probably the AGN feedback mechanism
\citep[e.g.,][]{SR98, King, Murray05, DSH05}, on a timescale $\la \taus$.

\item The majority of high-luminosity ($L_{\rm bol}\ga 10^{45.75}\ergs$) QSOs
accrete material via an almost single Eddington ratio, close to 1 and not
smaller than half of the Eddington limit, which suggests that the disk
accretion onto MBHs should be indeed self-regulated by the Eddington limit when
the accretion material is over-supplied in the initial phase.

\item Low-luminosity QSOs ($L_{\rm bol}\la 10^{45.25}\ergs$) accrete
material via a much wider range of Eddington ratios, and a significant fraction
of them accrete material via low Eddington ratio ($\dot{m}\la 0.1$), which
corresponds to the self-similar long-term evolution of disk accretion around
MBHs ($\bh\sim \tau^{-\gamma}$ and $\gamma\sim 1.2-1.3$) when the accretion
material is under-supplied.

\item The Eddington ratio distribution among QSOs/AGNs inferred from the
extended So{\l}tan argument concentrates toward high Eddington ratios (close to
1), especially for high luminosity QSOs, which appears to conflict with that
estimated directly from observations (with a mean value of $\sim
10^{-0.6}-10^{-1.1}$) by using the virial mass estimator(s). To make these two
distributions consistent with each other, an offset of $0.3-0.6$~dex in the MBH
mass estimated from the virial mass estimator(s) is required, at least for
high-luminosity QSOs, which suggests that MBHs masses obtained from the virial
mass estimator(s), have been systematically over-estimated by a factor of $2-4$.

\item The fraction of optically obscured QSOs/AGNs inferred from the extended
So{\l}tan argument can be as high as $80\%$ at $M_{\rm B}\sim -20$---$-23$ and
slightly decreases to $60\%$ at $M_{\rm B}=-24$---$-27$, and these numbers are
consistent with recent observations by \citet{Reyes08}. The dependence of the
fraction of type 2 AGNs on the X-ray luminosity cannot be solely an
evolutionary effect arising from their individual evolution after nuclear
activity is triggered (i.e., QSOs are more likely to be obscured in the early
stage of the MBH growth), and some other effects (e.g., those introduced by the
receding torus model; \citealt{Lawrence91}) should be responsible for the
larger fraction of type 2 AGNs at low luminosities ($L_{X}\la
10^{43.5}\ergs$).

\end{itemize}

We estimate possible effects due to MBH mergers (which may re-shape the local
BHMF) and multiple times of nuclear activity and accretion (e.g., triggered by
multiple times of galaxy `wet' major mergers) in the growth history of a
MBH, and we find that these effects on our conclusions are insignificant, which
again supports that the constraints obtained above are robust.

The constraints on the luminosity evolution of individual QSOs obtained from
the extended So{\l}tan argument in this paper, together with the QSO LF, can be
further used to derive the BHMF at high redshifts and the cosmic evolution of
the triggering rate of nuclear activity. These constraints and those recent
estimates on the Eddington ratio distribution in QSOs ask for serious
theoretical modeling of the long-term evolution of disk accretion around MBHs.
More detailed modeling of accretion and radiation transfer physics in the
vicinity of MBHs may have to be involved to determine the self-regulation of
the disk accretion (rather than the simple Eddington limit argument) at the
initial phase with sufficient deposited accretion material. It should be one of
the important long-term goals for theoretical studies on the growth of MBHs to
answer questions like what determines the transition from the initial rapid
accretion phase with Eddington ratio close to 1 to the rapid declining phase,
what shuts off the efficient accretion process around MBHs (probably
jointly determined by an efficient feedback mechanism and the accretion disk
viscosity), and what determines the evolution of the spin of MBHs. 

\acknowledgements{
We benefited from numerous discussions with Scott Tremaine, and we are indebted
to him for his contribution to this work, his comments on the drafts, and his
support during our visit to Institute for Advanced Study, where a significant
part of this work was done. We thank Rashid Sunyaev and Scott Tremaine for
their encouragements on using X-ray AGNs to constrain the growth of massive
black holes. We benefited from conversations with Nadia Zakamska on
Compton-thick AGNs.  This work is supported in part by NASA grants NNX08AH24G
and NNX08AL41G.
}


\end{document}